%% file: 0main.tex
\RequirePackage{fix-cm}
\documentclass[twocolumn]{svjour3}          
\smartqed  
\usepackage{graphicx}
\usepackage{xspace}
\usepackage{enumitem}
\usepackage{amssymb}
\usepackage[linesnumbered,ruled,vlined,boxed,noend]{algorithm2e}
\usepackage{booktabs}
\usepackage{multirow}
\usepackage{amsmath}
\usepackage{colortbl}
\usepackage{xcolor}

\newcommand{\myparagraph}[1]{\vspace{0.3\baselineskip}\noindent{\textbf{#1.}}~}
\newcommand{\finding}{\noindent{\underline{\textit{Observations}}}.~}

\newcommand{\task}{GCDIA\xspace}
\newcommand{\ti}{GCDI\xspace}
\newcommand{\ta}{GCDA\xspace}

\newcommand{\mmdb}{\texttt{GredoDB}\xspace}

\newcommand{\mmdbd}{\texttt{GredoDB-D}\xspace}
\newcommand{\mmdbs}{\texttt{GredoDB-S}\xspace}
\newcommand{\agens}{\texttt{AgensGraph}\xspace}
\newcommand{\arango}{\texttt{ArangoDB}\xspace}
\newcommand{\bench}{\texttt{M2Bench}\xspace}
\newcommand{\grain}{\texttt{GRainDB}\xspace}
\newcommand{\duck}{\texttt{DuckDB}\xspace}
\newcommand{\duckpgq}{\texttt{DuckPGQ}\xspace}
\newcommand{\poly}{\texttt{Polyglot}\xspace}
\newcommand{\grfusion}{\texttt{GRFusion}\xspace}

\newcommand{\ag}{adjacency graph\xspace}
\newcommand{\ra}[1]{\renewcommand{\arraystretch}{#1}}

\begin{document}

\title{Graph-centric Cross-model Data Integration and Analytics in a Unified Multi-model Database}




\author{Zepeng Liu$^{1}$  \and
        Sheng Wang$^{1}$  \and
        Shixun Huang$^{2}$\and
        Hailang Qiu$^{1}$ \and
        Yuwei Peng$^{1}$  \and
        Jiale Feng$^{1}$  \and
        Shunan Liao$^{1}$ \and
        Yushuai Ji$^{1}$  \and
        Zhiyong Peng$^{1,3}$
}


\institute{
    Zepeng Liu  \at
    \email{liuzp\_063@whu.edu.cn}
    \and
    Sheng Wang \at
    \email{swangcs@whu.edu.cn}
        \and
    Shixun Huang \at
    \email{shixunh@uow.edu.au}
        \and
    Hailang Qiu \at
    \email{helloqiu@whu.edu.cn}
    \and
    Yuwei Peng \at
        \email{ywpeng@whu.edu.cn}
    \and
    Jiale Feng \at
        \email{jlfeng@whu.edu.cn}
    \and
    Shunan Liao \at
        \email{snliao@whu.edu.cn}
    \and
    Yushuai Ji \at
        \email{yushuai@whu.edu.cn}
    \and
    Zhiyong Peng \at
        \email{peng@whu.edu.cn}
    \at
    $^{1}$ School of Computer Science, Wuhan University, Wuhan, China
    \at
	$^{2}$ School of Computing and Information Technology, The University of Wollongong, Wollongong, Australia
    \at
    $^{3}$ Big Data Institute, Wuhan University, Wuhan, China
}

\date{Received: date / Accepted: date}

\maketitle

\begin{abstract}
{
Graph-centric cross-model data integration and analytics (GCDIA) refer to tasks that leverage the graph model as a central paradigm to integrate relevant information across heterogeneous data models, such as relational and document, and subsequently perform complex analytics such as regression and similarity computation.
As modern applications generate increasingly diverse data and move beyond simple retrieval toward advanced analytical objectives (e.g., prediction and recommendation), GCDIA has become increasingly important.
Existing multi-model databases (MMDBs) struggle to efficiently support both integration (GCDI) and analytics (GCDA) in \task. They typically separate graph processing from other models without global optimization for GCDI, while relying on tuple-at-a-time execution for GCDA, leading to limited performance and scalability.}
To address these limitations, we propose \mmdb, a unified MMDB that natively supports storing \underline{g}raph, \underline{re}lational, and \underline{do}cument models, while efficiently processing GCDIA. Specifically, we design 1) topology- and attribute-aware graph operators for efficient predicate-aware traversal, 2) a unified GCDI optimization framework to exploit cross-model correlations, and 3) a parallel GCDA architecture that materializes intermediate results for operator-level execution.
Experiments on the widely adopted multi-model benchmark \bench demonstrate that, in terms of response time, \mmdb achieves up to $107.89 \times$ and an average of $10.89 \times$ speedup on GCDI, and up to $356.72 \times$ and an average of $37.79 \times$ on GCDA, compared to state-of-the-art (SOTA) MMDBs.
\end{abstract}

\input{1intro}
\input{7Preliminary}
\input{2RelatedWork}

\input{4Algorithms}
\input{3Definitions}
\input{5Optimizations}

\input{6Experiments}

\section{CONCLUSIONS AND FUTURE WORK}
In this work, we presented \mmdb, a unified multi-model database that natively supports relational, document, and graph models within a single system, to address long-standing challenges in efficiently executing \task.
Specifically, to achieve efficient \ti, we introduce a novel operator-based framework that integrates newly designed graph-centric and cross-model operators, together with the first \ti optimization strategies for efficient planning.
To overcome the functional and performance limitations of volcano model-based \ta processing, \mmdb integrated a new parallel analytical pipeline, featuring dedicated matrix generation, intermediate result reuse, and parallel analytical operators.
Experimental results on the multi-model benchmark demonstrated that \mmdb significantly outperformed existing MMDBs on \task workloads.

In future work, we plan to extend \mmdb into a distributed MMDB and incorporate advanced query optimization techniques, such as learning-based cost estimation and planning. Moreover, we will explore large language model-based methods to improve usability and lower the learning curve for interacting with MMDBs.



\begin{acknowledgements}
This work was supported by the National Key R\&D Program of China (2023YFB4503600), National Natural Science Foundation of China (62202338), and the Key R\&D Program of Hubei Province (2023BAB081).
\end{acknowledgements}

\bibliographystyle{spmpsci}      
\bibliography{sample-base}   


\end{document}

%% file: 1intro.tex
\section{INTRODUCTION}
\label{sec: intro}

\begin{figure*}[t]
    \centering
    \includegraphics[width=\textwidth]{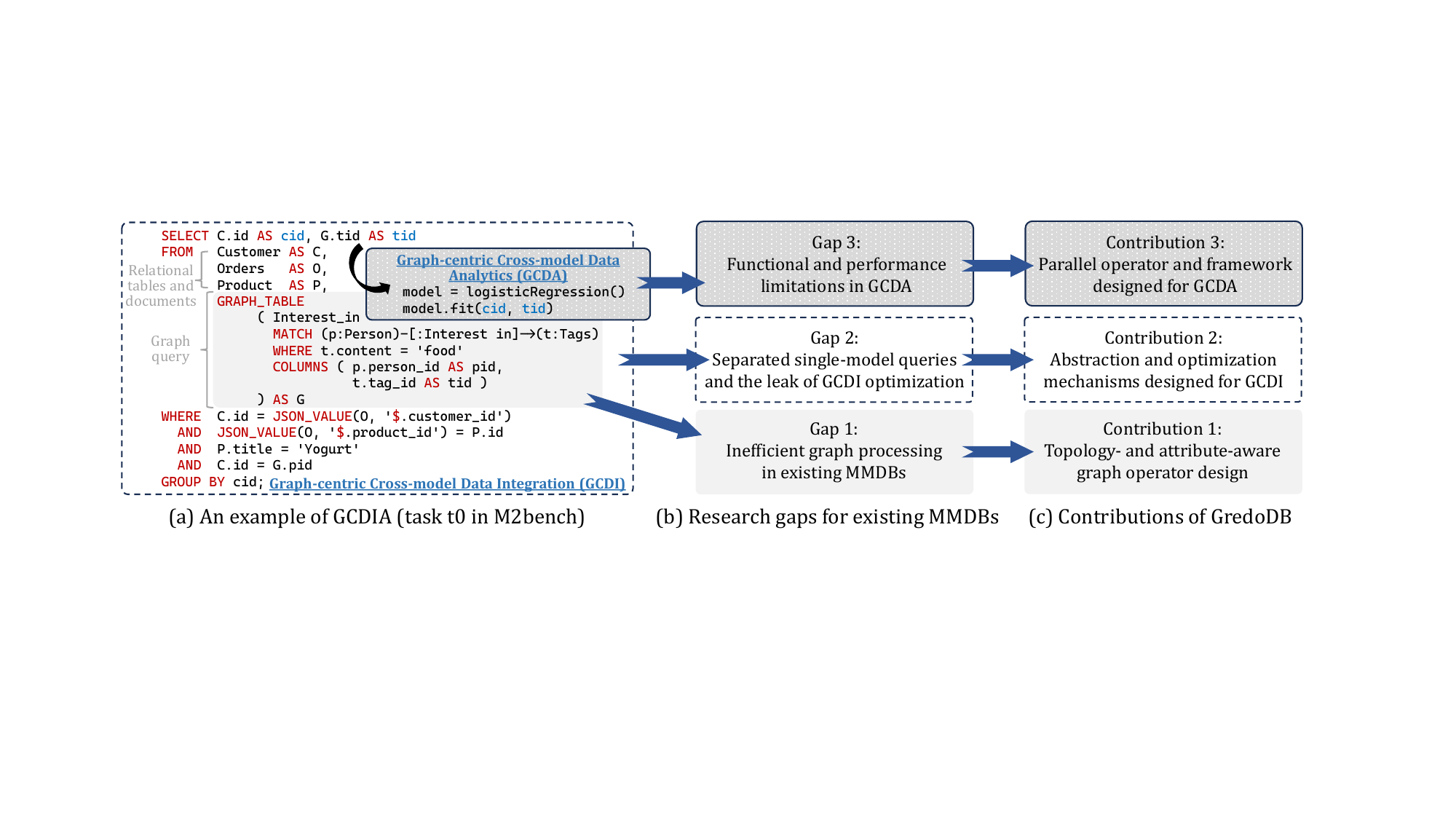}
    \caption{Overview of research problems, gaps, and how \mmdb addresses them. In (a), the query is graph-centric, as it uses the graph topology as the primary means to connect entities represented by the \texttt{Customer} (cid) and \texttt{Tag} (tid) vertices.} 
    \label{fig: intro gcq}
\end{figure*}

Modern data-driven applications, such as e-commerce platforms \cite{kellyton2021social}, knowledge graph-based search engines \cite{hogan2021knowledge,SmartBench2022Orogat}, and social media services \cite{Patterson2010Interaction,Mendoza2010Twitter}, increasingly rely on advanced numerical analytic, including logistic regression and cosine similarity, to support tasks like user profiling \cite{Yin2014user,Yin2015Dynamic,zhu2023f3km} and recommendation \cite{Chandramouli2011StreamRec,wu2022survey}. 
With the increasing diversity of data sources, the data required for such analytics is often distributed across heterogeneous data models, including structured relational tables, semi-structured documents, and unstructured graphs. For instance, considering the following example in \bench \cite{kim2022m2bench}, a benchmark designed for multi-model data integration and analytics:

\textit{In an e-commerce scenario, the data platform stores vast amounts of information about its customers, products, and the interests of each customer in their original structure. Specifically, these heterogeneous data contain: 1) two relational tables \texttt{Product(id,title,price)} and \texttt{Customer(id,person\_id,name)}; 2) a document collection \texttt{Orders} with keys \texttt{customer\_id} and \texttt{product\_id}; and 3) an \texttt{Interested\_in} graph consisting of vertices \texttt{Person} and \texttt{Tags} with edges labeled as \texttt{Interested in}. To understand user preferences, data analysts may pose analytics over these multi-model data, such as: ``build a logistic regression model to predict which food-related tags are followed by users who have purchased yogurt", which requires aggregating the required entities across different models, and then feeding the query results into downstream analytical tools to complete the analysis.}

\myparagraph{The Research Problem We Study}
{In this example, the overall task considered by the data analysts is referred to as \underline{Graph-centric Cross-model Data Integra-} \underline{tion and Analytics (\task)}.
We formally define \task as tasks that integrate target data from heterogeneous data models
and perform analytical processing over the integrated data.
In a \task, the graph model serves as the primary query structure for expressing relationships
and traversal semantics, while other models such as relational and document provide complementary attributes
and analytical features.                
}
Although the term \task is relatively new, the fundamental problem it addresses has long been prevalent in practice \cite{duggan2015bigdawg,joining2024fu,Jezek2025DortDB,kim2022m2bench}. 
Figure~\ref{fig: intro gcq}(a) illustrates the concrete queries and operations involved in this process, following the SQL/PGQ standard \cite{deutsch2022graph} for graph querying and document path expressions \cite{durner2021json} for accessing document data.
{Conventionally, \task can be decomposed into two subtasks, namely graph-centric cross-model data integration (\ti) and graph-centric cross-model data analytics (\ta). The latter is performed on top of the results produced by \ti, while \ti can exist as an independent task.}
Motivated by its practical significance,
this paper focuses on efficiently enabling the execution of \task.

\myparagraph{Research Gaps}
{A naive solution to support \task is to build multi-engine systems (MESs) }\cite{Michael2015Spark,boyan2016cloud}, which store different data models in separate databases, and leverage a specific analytical engine for \ta. However, as large data volumes lead to severe data movement overhead between engines, MESs become inefficient for \task shown in Figure~\ref{fig: intro gcq}(a).
Another promising solution is to use multi-model databases (MMDBs) \cite{ritter2021orientdb,lu2019mmdb}. MMDBs integrate multiple data models within a single database, thereby natively eliminating data movement overhead in \ti.
By supporting multiple data models within a single system, MMDBs offer significant potential for cross-model query processing and optimization. For example, systems such as \arango \cite{ArangoDB2025} and \agens \cite{agensgraph} support integrated querying over relational and document models, allowing limited forms of dual-model query optimization. Despite these advantages, in graph-centric scenarios, efficiently supporting \ti and \ta remains a major challenge. This limitation stems from fundamental architectural issues in current MMDB designs, as shown in Figure~\ref{fig: intro gcq}(b). We summarize the key research gaps as follows:

\begin{itemize}[left=0pt]
    \item \textit{Topology- or attribute-agnostic graph queries.}
    Many existing MMDBs \cite{agensgraph} translate graph queries into join-heavy plans over their primary data model, which fails to exploit graph topology and leads to inefficient execution; {we refer to them as translation-based systems (TBSs)}. In contrast, some systems \cite{Hassan2018Extending,MongoDB2025MongoDB} natively support graph storage and provide traversal-oriented operators (e.g., DFS or BFS), which excel at exploring graph topology but largely ignore attribute semantics; {we refer to them as graph-native systems (GNSs)}. Consequently, current approaches tend to be either topology-agnostic or attribute-agnostic, making them ill-suited for \ti that tightly integrate graph structure with multi-model attributes.

    \item \textit{Leaking global optimization for \ti.}
    MMDBs such as \agens typically employ logically isolated processing pipelines for different data models \cite{lee2024chimera,Mhedhbi2022Modern}. This fragmentation prevents global planning and optimization, forcing \ti to be decomposed into disjoint single-model subqueries executed independently. Consequently, \ti cannot exploit cross-model correlations (e.g., jointly pruning candidate records via multi-model predicates, or generating unified \ti plans), leading to sub-optimal execution efficiency.

    \item \textit{Tuple-at-a-time analytics processing.}
    When processing batch or iterative \ta, such as regression, existing MMDBs largely rely on the volcano model’s tuple-at-a-time execution paradigm \cite{graefe1993volcano}. This design incurs high execution overhead due to excessive iterator invocations, function call overheads, and poor cache locality \cite{crotty2015architecture}. 
    Some studies \cite{wang2017myria} introduce an additional analytical engine into MMDBs to support \ta, but suffer from the same shortcomings as MESs, as this design incurs additional data movement overhead.

\end{itemize}
Addressing these gaps remains challenging in practice. At the query planning level, most MMDBs lack a unified abstraction for interactions among graph, relational, and document data. As a result, cross-model semantics and optimization rules are largely missing, preventing global \ti planning and optimization. Furthermore, supporting \ta requires execution engines that can natively and efficiently handle iterative and compute-intensive operations, while avoiding the data movement overhead incurred by introducing external engines.

\myparagraph{Our Methodology and Contributions}
In this paper, we design \mmdb, an advanced MMDB that \emph{unifies the management of graph, relational, and document data within a single system, while efficiently supports \task}. To tackle the aforementioned fundamental and long-standing challenges, we contribute three core innovations in \mmdb, as shown in Figure~\ref{fig: intro gcq}(c):
1) \textit{Topology- and attribute-aware graph operator design.} We design a new class of graph query operators that jointly exploit graph topology and attribute predicates during execution. Unlike TBSs that ignore graph topology, or GNSs that treat attribute filtering as a secondary concern, our operators enable predicate-aware traversal. This enables more efficient and selective graph traversals, reducing unnecessary exploration of irrelevant records and topology nodes.
2) \textit{\ti abstraction and optimization framework.}
We introduce a unified \ti abstraction that treats graph, relational, and document models as first-class citizens, each with dedicated logical and physical operators. This abstraction enables \mmdb to explicitly capture cross-model interactions and to generate global cross-model query plans, rather than decomposing \ti into isolated subqueries. Built upon this abstraction, we develop a systematic \ti optimization framework that exploits predicate- and record-level correlations across models. The framework enables principled cross-model optimizations, such as predicate propagation, candidate filtering, and sub-plan pushdown, thereby significantly improving the efficiency of \ti execution.
3) \textit{Parallel analytic architecture}. We integrate \ta into the query execution pipeline by designing analytical operators and integrating them into the execution plan. \mmdb materializes \ti results into an intermediate analytical storage layer, which serves as a shared input for downstream analysis operators. By allowing analytical operators to participate directly in query planning and execution, \mmdb enables operator-level parallelism and data reuse across heterogeneous processing stages. This design avoids tuple-at-a-time execution and data movement overhead,
enables efficient, parallel \ta processing.
{Overall, we summarize the capabilities of \mmdb and existing systems in Table~\ref{tab: related work} and highlight our key contributions below:}

\begin{itemize}[left=0pt]
    \item We design and implement \mmdb, a unified MMDB that supports \ti over relational, document, and graph models, as well as efficient \ta processing.
    \item We propose topology- and attribute-aware operators for efficient graph processing, enabling predicate-aware traversal and flexible cost-based planning.
    \item We present a unified \ti abstraction and optimization framework, with novel planning and optimization rules that enable tightly integrated, correlation-aware multi-model execution.
    \item We integrate \ta into the query execution pipeline via operator-driven parallel analysis, enabling efficient in-database \ta processing.
    \item Experiments on \bench show that \mmdb achieves over $15.02 \times$ and $63.23 \times$ average speedups on \ti and \ta, respectively, compared to SOTA MMDBs.
\end{itemize}

\input{related_work}

%% file: related_work.tex
\begin{table}[t]
\setlength{\tabcolsep}{2.2pt}
\centering
\ra{1.03}
\caption{System capabilities across architectures for \task.}
\label{tab: related work}
\begin{tabular}{ccccc}
\toprule




\multirow{2}{*}{\textbf{\rule{0pt}{3.5ex}Features}} & \multirow{2}{*}{\textbf{\rule{0pt}{3.5ex}MESs}} & \multicolumn{3}{c}{\textbf{MMDBs}} \\ \cmidrule{3-5}

&  & {\textbf{TBSs}} & {\textbf{GNSs}} & \mmdb \\ \midrule

Graph-native Storage   & \checkmark    
& $\times$ 
& \checkmark    & \checkmark       \\

Topology-aware Processing   & \checkmark    
& $\times$ 
& \checkmark     & \checkmark       \\

Attribute-aware Processing   & \checkmark    
& \checkmark 
& $\times$     & \checkmark       \\

Translating Overhead-free & $\times$  
& $\times$  
& \checkmark    & \checkmark        \\
        
\ti Optimization  & $\times$
& $\times$  
& $\times$    &   \checkmark  \\

Dedicated \ta Support  & \checkmark 
& $\times$  
& $\times$  &  \checkmark   \\

No Data Movement  & $\times$ 
& \checkmark  
& \checkmark  &  \checkmark   \\

\bottomrule
\end{tabular}

\end{table}

%% file: 7Preliminary.tex
\section{RELATED WORK}\label{sec: related work}
In this section, we review existing systems from an architectural perspective, focusing on how they support \task. We organize prior work based on their query processing architectures, and highlight the fundamental limitations and research gaps of these designs.

\subsection{Multi-engine Systems}
MESs, including polyglot systems \cite{Michael2015Spark,survey2022Ali,Alekh2014VERTEXICA}, polystore systems \cite{ben2019answering,duggan2015bigdawg,wang2017myria,shi2024font,Lei2024Xstor}, and multistore systems \cite{boyan2016cloud,daniel2017neoemf}, employ multiple single-model databases to manage heterogeneous data models independently, and rely on a dedicated analytical engine to process \ta.
When processing \ti, MESs typically decompose the \ti into several single-model subqueries, each of which is dispatched to a dedicated engine specialized for a particular model \cite{duggan2015bigdawg,Lei2024Xstor}. After execution, each engine transfers the intermediate results to another engine for further processing, or to a centralized coordination layer, where the intermediate results are combined to generate the final \ti output.

While such an architecture enables native storage and execution for individual data models and can be extended with specialized engines to support \ta, it introduces fundamental limitations for \task. Specifically, intermediate results generated by different subqueries must be materialized and transferred across engines, leading to substantial communication and data movement overhead \cite{duggan2015bigdawg,kim2022m2bench}. Moreover, cross-engine execution enforces strict isolation among query processing stages of different models, which prevents joint optimization of \ti. Prior studies \cite{lu2019mmdb} have also reported that MESs suffer from significant data redundancy and complex consistency management across engines. These architectural drawbacks prevent MESs from efficiently supporting the \task task considered in this work.

\subsection{Multi-model Databases}
We categorize MMDBs as TBSs and GNSs based on their support strategies for the graph model.

\myparagraph{Translation-based Systems}
TBSs \cite{agensgraph,duckdb} translate graph model into their primary storage model, such as the relational model in \duck \cite{duckdb}, and implement graph queries via index-accelerated multi-way joins. By compiling graph constructs into equivalent operations over the underlying model, TBSs can reuse existing storage layouts, execution engines, and optimization frameworks to support graph queries without introducing dedicated graph-oriented physical operators. For example, \agens \cite{agensgraph} translates pattern matching operations in a \ti into B-tree accelerated multi-way joins under its relational architecture.

Despite enabling basic support for \ti, TBSs exhibit both functional and performance limitations when processing \ti: 1) TBSs can only support a small class of graph queries, as some graph semantics (e.g., shortest-path search) cannot be faithfully expressed or efficiently implemented using physical operators originally designed for the primary data model; and 2) due to the lack of native access to graph topology, pattern matching operations in TBSs are typically implemented using sequences of equality joins and self-joins over edge representations. When processing multi-hop \ti, this multi-way joins execution strategy often leads to an exponential growth of intermediate results, severely degrading performance \cite{leis2015job,sahu2020ubiquity}. Moreover, constrained by tuple-at-a-time execution models inherited from traditional database architectures, TBSs are generally limited to supporting simple numerical aggregations. As a result, they are inefficient, or even incapable, of executing computation-intensive \ta tasks that require batch-oriented or matrix-based processing.

\myparagraph{Graph-native Systems}
Prior studies \cite{Skavantzos2025Entity,Skavantzos2025ER} show that graph-native processors can significantly outperform join-based relational execution for graph queries. This observation has motivated much research on GNSs \cite{lee2024chimera,luo2025relgo,jin2022graindb,jin2022making,Hassan2018Extending,lin2016fast,ArangoDB2025}, which aim to extend relational or document database systems with native graph storage and query processing capabilities. To preserve graph topology within a non-graph-native engine, GNSs typically adopt adjacency lists \cite{lee2024chimera,luo2025relgo,Hassan2018Extending} or predefined join indexes \cite{jin2022graindb,jin2022making,ArangoDB2025}, and introduce specialized physical operators to execute graph queries efficiently. In particular, many GNSs implement traversal-oriented operators based on DFS or BFS execution \cite{Hassan2018Extending}, which significantly reduce the cost of pattern matching by avoiding expensive multi-way joins over relational representations.

Although existing GNSs perform effectively on pure graph workloads, they exhibit notable limitations in the context of \task. On the \ti side, their traversal operators are primarily designed for topology-oriented execution and have limited awareness of attribute predicates and record-level semantics. Moreover, most GNSs enforce a clear logical separation between graph and non-graph models, as well as between their corresponding processing pipelines, which restricts cross-model interaction and reduces optimization opportunities for \ti workloads. On the \ta side, GNSs either suffer from the same efficiency issues as TBSs
due to tuple-at-a-time execution, or, similar to MESs, rely on additional analytical engines that introduce extra execution overhead.

\subsection{Systems for Analytical Workloads} \label{sec: Systems for Analytical Workloads}
A large body of prior work in both academia \cite{duckdb,Jin2023kuzu} and industry \cite{Padmanabhan2001ibm,Boncz2005monetdb} has focused on enabling database management systems to efficiently process analytical workloads since the early 2000s. Representative systems such as IBM DB2 \cite{Padmanabhan2001ibm} and MonetDB/X100 \cite{Boncz2005monetdb} introduced vectorized execution, late materialization, and cache-conscious processing to exploit intra-query parallelism and modern CPU architectures for analytical queries. More recently, \duck further advances this line of work by adopting a fully columnar storage layout and a vectorized execution engine optimized for scan- and aggregation-heavy analytical workloads.

However, these techniques primarily target traditional relational analytics, especially aggregation-centric query fragments involving GROUP BY and HAVING clauses. Their execution models and optimization strategies are fundamentally designed around relational operators like aggregation. When facing \ta such as regression, these systems often suffer from both functional limitations and performance bottlenecks due to architectural constraints, including rigid operator pipelines and limited support for iterative computation patterns.

%% file: 2RelatedWork.tex
\begin{figure*}
    \centering
    \includegraphics[width=\textwidth]{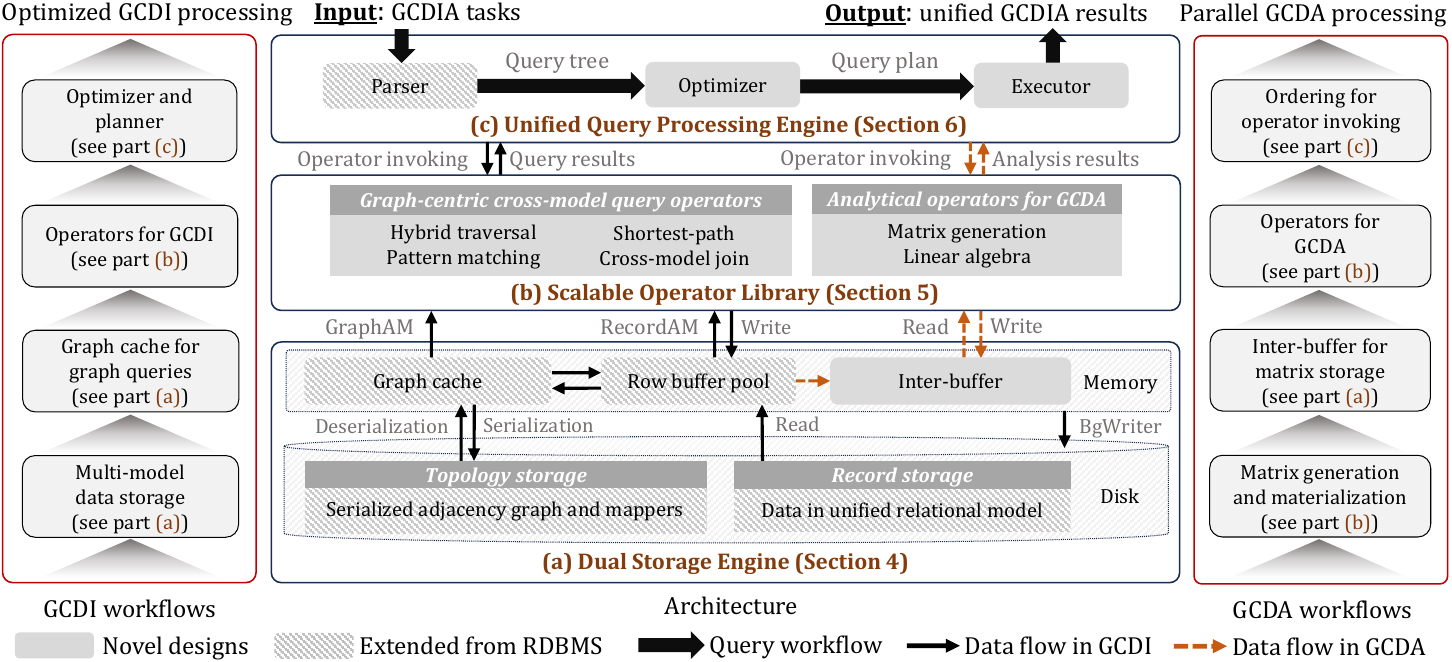}
    \caption{An overview of the architecture of \mmdb. The bold black arrows indicate the primary execution pipeline, while the thin black and brown arrows respectively denote data flows in \ti and \ta processing.
    }
    \label{fig:overview}
\end{figure*}

{
\section{PRELIMINARIES}
In this section, we introduce the notations used in this paper and present the formulation of \task. We then provide an overview of \mmdb, illustrating how its components collaborate to efficiently support \task.

\subsection{Notations}

\input{notations}

Table~\ref{tab: notations} summarizes the main operators used in this paper. 
Among them, the hybrid traversal operator $\mapsto$ and the graph pattern operator $\mathcal{P}$ are newly designed to enable topology- and attribute-aware graph queries. 
The cross-model join operator $\widehat{\bowtie}$ formalizes interactions across multiple data models, serving as a foundation for \ti optimization. 
The remaining symbols are included for completeness in \task formulation.}

\subsection{\task Formulation}
Existing MMDBs typically adopt translation-based (e.g., TBSs) or loosely separated processing strategies (e.g., GNSs) when handling multi-model data. A common limitation of these approaches is the lack of a unified abstraction and explicit interaction rules for \task, which fundamentally restricts opportunities for global planning and optimization of both \ti and \ta. In contrast, we introduce the \emph{first unified abstraction for \task} here.

\myparagraph{\ti Formulation}
We first formulate the \ti part, which integrates entities across graph-centric multi-model data, as a \texttt{Select}-\texttt{From}-\texttt{Match}-\texttt{Where} (SFMW) task. This formulation extends the traditional relational query model with an explicit \texttt{Match} clause for graph pattern specification, and can be expressed as follows:
\begin{equation} \label{eq: sfmw}
T_{\ti} = \pi_A(\sigma_\Psi(H_1 \widehat{\bowtie}_{\mathcal{F}_1} \dots \widehat{\bowtie}_{\mathcal{F}_{k-1}} (\widehat{\pi}_{A_k'} \mathcal{P}(H_k, P_k)))).
\end{equation}
The formulation in Eq.~\eqref{eq: sfmw} involves $k$ data collections, each $H \in \{R, D, G\}$. $\pi_A$ denotes a projection over attributes $A=\{a_1, \dots, a_m\}$, where each $a_i \in A$ may refer to a relational attribute, a document path expression, or a property of a graph vertex or edge. Similarly, $\sigma_{\Psi}$ denotes a selection with a set of predicates $\Psi$. Note that each graph is consistently associated with a pattern matching operation $\mathcal{P}$, followed by a {graph projection} $\widehat{\pi}_{A'}$ over the graph-relation produced by $\mathcal{P}$, where $A'$ denotes the set of columns to be selected.
Based on the above formulation, the example described in Section~\ref{sec: intro} can then be represented as follows:
\begin{gather}
T_1 = \texttt{C} \widehat{\bowtie}_{\texttt{C}.\texttt{id=p}.\texttt{id}} (\widehat{\pi}_{\texttt{p}.\texttt{id,t}.\texttt{id}} \mathcal{P}(\texttt{Interested in}, P)), \label{eq: q1} \\ 
T_2 = \texttt{P} \widehat{\bowtie}_{\texttt{P}.\texttt{id=O->}\texttt{>`product\_id'}} \texttt{O} \widehat{\bowtie}_{\texttt{O->}\texttt{>`customer\_id'=C}.\texttt{id}} T_1, \label{eq: q2} \\
T_3   = \pi_{\texttt{C}.\texttt{id,t}.\texttt{id}}(\sigma_{\texttt{P}.\texttt{title="Yogurt"}} T_2), \label{eq: q3}
\end{gather}
where the pattern $P$ in Eq.~\eqref{eq: q1} has the structure defined in Figure~\ref{fig: intro gcq}(a).

\myparagraph{\ta and \task Formulation}
A \ta task performs complex analytical processing, such as matrix multiplication
and matrix similarity computation, over the results produced by \ti.
In \mmdb, we first transform the output of \ti into a matrix representation
via a function $\mathcal{G}$, and then apply analytical processing on this matrix
using a function $\mathcal{A}(\cdot)$.
Accordingly, a \ta task can be formulated as:
\begin{equation}
T_{\ta} = \mathcal{A}(\mathcal{G}(\cdot)).
\end{equation}
Base on this, we formulate \task as:
\begin{equation}
T_{\task} = \mathcal{A}(\mathcal{G}(T_{\ti})).
\end{equation}

\subsection{System Overview}
\label{sec: overview}

Figure~\ref{fig:overview} illustrates how \mmdb supports \task in a unified architecture.
We organize this paper by introducing the main components: the dual storage engine (Section~\ref{sec: storage}), the scalable operator library (Section~\ref{sec: operators}), and the unified query processing engine (Section~\ref{sec: execution engine}).

\myparagraph{Dual Storage Engine}
As shown in Figure~\ref{fig:overview}(a), to support relational, document, and graph data within a unified system, \mmdb adopts a dual storage engine with catalog extensions for heterogeneous schema representation. The storage layer consists of: 1) a unified \textit{record storage} that stores records across all data models, and 2) a dedicated \textit{topology storage} that preserves graph topology in the form of an \textit{adjacency graph}, enabling efficient graph traversal.

The idea of preserving graph topology using adjacency structures is not new and has been widely adopted by GNSs \cite{Hassan2018Extending,luo2025relgo,jin2022graindb}. The key novelty of \mmdb lies in establishing a precise one-to-one mapping between records in the unified record storage and vertices in adjacency graph, and \emph{encapsulating this mapping as a dedicated physical operator}. This design enables seamless interaction between graph topology and attribute data during query execution, addressing the inefficiencies of topology-agnostic joins, as well as pure topology-driven traversal that ignores the attribute filters. As a result, \mmdb can jointly capture and optimize both structural and attribute semantics in \ti.

To support \ta, \mmdb further introduces an in-memory inter-buffer that materializes intermediate results produced by \ti. By caching \ti results at the record-batch level, the inter-buffer eliminates the need for tuple-at-a-time data production and allows analytical operators to directly consume these intermediate results as inputs. This design enables efficient data reuse, supports operator-level parallelism, and removes the data movement overhead of \task.

\myparagraph{Scalable Operator Library}
Many existing systems attempt to extend functionality with minimal engineering effort by reusing their original execution frameworks. For example, \texttt{SQLGraph} \cite{sun2015sqlgraph} and \duckpgq \cite{duckpgq} support graph queries by translating them into relational operations and executing them within an existing relational engine. While this translation-based approach lowers implementation cost, it often leads to fundamental limitations in both functionality and performance. To address these issues, \mmdb introduces a scalable operator library that is tightly coupled with its dual storage engine, as illustrated in Figure~\ref{fig:overview}(b). Rather than relying on query translation, \mmdb designs dedicated physical operators for both \ti and \ta.

For \ti, we provide physical operators that jointly perform topology exploration and attribute predicate evaluation. These operators directly exploit the adjacency graph for efficient traversal while enabling fine-grained interaction with vertex and edge records. In addition, \mmdb introduces a dedicated cross-model join operator that unifies records from heterogeneous models, allowing seamless data integration in \ti.

For \ta, \mmdb offers a parallel analytical operator library that consumes materialized \ti results. These operators enable \ta such as logistic regression and similarity computation, avoiding tuple-at-a-time processing and external data movement. All operators expose standardized interfaces to the storage layer, ensuring efficient execution and seamless integration with the planner and execution engine.

{
\myparagraph{Unified Query Processing Engine for \task}
As shown in Figure~\ref{fig:overview}(c), \mmdb provides a unified query processing engine for \task. User queries are expressed in an SQL/PGQ-compatible language.

First, for \ti workflows (the left part of Figure~\ref{fig:overview}), when a \ti is issued, the relevant multi-model data is loaded from disk into memory. In particular, the graph topology is deserialized into an adjacency graph and cached in the graph cache, enabling efficient access by the hybrid traversal operator. Based on a unified abstraction, \mmdb employs a dedicated planner
to compose physical operators into executable \ti plans.
A dedicated cost model further enables correlation-aware optimization strategies that go beyond repurposed relational planning techniques.

Second, for \ta workflows (the right part of Figure~\ref{fig:overview}), \mmdb first materializes the \ti results into matrix representations and places them in an in-memory inter-buffer. Based on user-specified analytical instructions, the system then constructs a plan by ordering the corresponding physical operators and submits it to the execution engine for evaluation.
In this design, \ta operators are typically placed at the upper levels of the query plan and operate on the results produced by \ti. By modeling \ta as plan-driven operator execution, \mmdb enables operator-level parallelism and supports efficient reuse of intermediate results across heterogeneous execution flows.}

\subsection{Summary and Clarification}
To support efficient \task execution, \mmdb introduces a set of specialized components that jointly address \ti and \ta\footnote{This paper focuses on the design and optimization of \mmdb for \task. The optimization of conventional relational, document query processing, and aggregation-centric analytical queries falls outside the scope of this work and is left for future research.}. For \ti, the design comprises: 1) a hybrid traversal operator, 2) a topology- and attribute-aware pattern matching operator, and 3) a \ti optimization framework. For \ta, \mmdb incorporates: 1) an inter-buffer for caching intermediate results, 2) a matrix generation method, and 3) a parallel operator-based analytical framework.

%% file: notations.tex
\begin{table}[t] \small
\setlength{\tabcolsep}{6pt}
\ra{1.0}
\centering
\caption{Summary of notations.}
\label{tab: notations}
\begin{tabular}{cc}
\toprule
\textbf{Symbol}     & \textbf{Description}                  \\ \midrule
$T$                 & The subtask or subquery    \\
$A$                 & The attribute set    \\
$R,D,G$             & Collection of relational, document, and graph        \\
$P$                 & The graph pattern \\
$\mathcal{F}$       & The predicate \\
$\mapsto$           & The hybrid traversal operator \\
$\mathcal{P}(G,P)$  & The pattern matching operation on $G$ \\
$\widehat{\bowtie}$ & The cross-model join operator\\
$\Psi$              & The set of predicates \\
$\pi_A$             & The projection over the attribute set $A$ \\
$\widehat\pi_A$     & The graph-projection over the attribute set $A$ \\
$\sigma_{\Psi}$     & The selection with a predicate set $\Psi$ \\
$\mathcal{G}$       & The matrix generation function \\
$\mathcal{A}$       & The abstraction of analytical tasks \\


\bottomrule
\end{tabular}
\end{table}

%% file: 4Algorithms.tex
\section{DUAL STORAGE ENGINE}
\label{sec: storage}

\subsection{Data Modeling}
\label{sec: Unified Disk Storage Model}

\begin{definition}[\textbf{{Relational Model}}]
A relation $R$ consists of a collection of tuples defined over a fixed schema $\{a_1, a_2, \dots, a_m\}$, in which each attribute $a_i$ corresponds to a column in the table.
\end{definition}

\begin{definition}[\textbf{{Document Model}}]
A document $d$ is a hierarchical collection of key-value pairs in JSON\footnote{In this paper, the document model refers to data represented in JSON format, as it is the most prevalent document representation in modern database systems. Other hierarchical document formats, such as XML, are also compatible with the proposed abstraction and can be supported in a similar manner.} format, where values may be atomic types (e.g., number, string) or nested documents/arrays. A document dataset $D = \{ d_1, d_2, \dots, d_n \}$ is modeled as an unordered set of such documents, typically sharing similar key structures and nesting patterns.
\end{definition}

\begin{definition}[\textbf{{Graph Model}}]
A graph is defined as $G = (\Omega, V, E, \mathcal{L})$, where $\Omega$ denotes the graph topology, stored in the form of an \textit{adjacency graph} (details discussed later); $V$ and $E$ are collections of vertex and edge records, respectively; and $\mathcal{L}(\cdot)$ is a labeling function assigning labels to vertices and edges.
\end{definition}

In \mmdb, each graph enforces a uniform edge label, i.e., for any $e_i, e_j \in E, \ \mathcal{L}(e_i) = \mathcal{L}(e_j)$, indicating a consistent semantic role for all edges. For instance, the network shown in Figure~\ref{fig: graph schema}(a) can be decomposed into two graphs: an \texttt{Interested\_in} graph with vertices $\{v_0, v_1, \dots, v_5\}$ and edges labeled \textit{Interested in}, and a \texttt{Follows} graph with vertices $\{v_1, v_2, v_4\}$ and edges labeled \textit{Follows}.

\begin{definition}[\textbf{{Adjacency Graph}}]
We define an adjacency graph $\Omega = (N_s, N_t, \mathcal{I})$ as a list-based representation of the topology of a graph, where $N_s$ and $N_t$ denote the sets of source nodes and target nodes, and $\mathcal{I}(\cdot)$ assigns a node identifier $nid$ to each node $n \in N_s \cup N_t$.
\end{definition}

Given a graph $G=(\Omega, V, E, \mathcal{L})$, the adjacency graph $\Omega$ comprises a source node set $N_s$ and a target node set $N_t$, where $|N_s| = |V|$\footnote{We use $|\cdot|$ to obtain the size of a set.} and $|N_t| = |E|$. Each node $n \in N_s \cup N_t$ contains a unique identifier $\mathcal{I}(n) = nid$ and a \textit{next} pointer. For any source node $n_s \in N_s$, the \textit{next} pointer references the first target node $n_t \in N_t$ in its adjacency list (or \texttt{Null} if no out-edges exist). Each target node, in turn, points to the next target node in the list (or \texttt{Null} if it is the last), forming a singly linked list that encodes the out-edges of the corresponding vertex in $G$. To support undirected graph queries and expose additional optimization opportunities, \mmdb maintains both forward and reverse adjacency graphs.
Notably, although \ag shares the same structure as the join-index used in some GNSs such as \grain \cite{jin2022graindb}, their usage is largely different. \grain leverages the predefined join-index to accelerate pure graph traversal, whereas \mmdb employs \ag in conjunction with record attributes, enabling topology- and attribute-aware graph processing. This design, in turn, allows \mmdb to apply the \ti optimization techniques introduced in Section~\ref{sec: execution engine}.

\subsection{Data Storage Format}
\label{sec: Data Storage Format}

\myparagraph{Unified Record Storage}
In \mmdb, multi-model data records, including tuples, documents, and graph vertex/edge records, are stored in the unified record storage (see Figure~\ref{fig:overview}(a)) within a unified relational format that permits non-first-normal-form (NF$^2$) \cite{jaeschke1982remarks,linnemann1987non} structures, allowing nested and multi-valued attributes to naturally represent document-style and array data. This unified layout enables consistent access to multi-model data records.

\begin{figure}[t]
    \centering
    \includegraphics[width=0.48\textwidth]{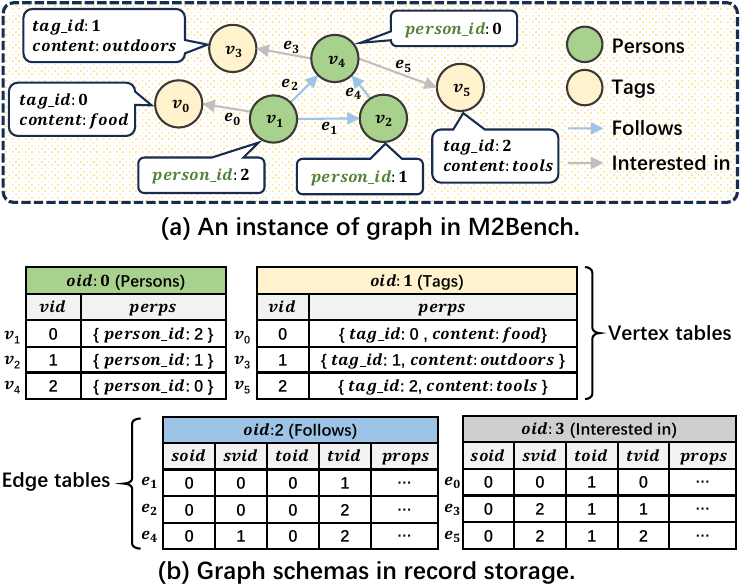}
    \caption{An example of graph storage format.}
    \label{fig: graph schema}
\end{figure}

For document data, similar to databases such as PostgreSQL \cite{PostgreSQL}, \mmdb utilizes the JSONB \cite{durner2021json} data type extension for storage, treating documents as special fields within relational tables. For a graph $G = (\Omega, V, E, \mathcal{L})$, \mmdb encodes each vertex and edge as a record comprising predefined structural keys, e.g., vertex ID ($vid$) and source/target IDs ($svid$/$tvid$), and a JSONB field that encapsulates its properties.
For instance, Figure~\ref{fig: graph schema} shows an example of graph storage format in the unified record storage, which corresponds to the property graphs \texttt{Follows} and \texttt{Interested\_in}. Each table is uniquely identified by an $oid$ and contains homogeneous vertices or edges sharing the same label. 
In vertex tables, each vertex is assigned a local identifier $vid$, along with its associated properties. Vertices are uniquely identified by the composite key $(oid, vid)$. For example, $(0, 0)$ refers to $v_1$. In edge tables, each edge is represented by its source and target vertices specified as $(soid, svid)$ and $(toid, tvid)$, respectively, and we leverage the composite key $(oid, soid, svid, toid, tvid)$ to uniquely identify each edge. We define the key retrieval methods for vertices and edges as follows:
\begin{itemize}[left=0pt]
    \item \textit{getVertexKey}: get the key of a vertex record $v \in V$. \\
    input = $v$; output = $(oid, vid)$.
    
    \item \textit{getEdgeKey}: get the key of an edge record $e \in E$. \\
    input = $e$; output = $(oid, soid, svid, toid, tvid)$.
\end{itemize}

\myparagraph{Graph-centric Topology Storage}
We store adjacency graphs in the topology storage in a serialized format (see Figure~\ref{fig:overview}(a)) for efficient in-memory reconstruction. During graph query execution, relevant adjacency graphs are selectively deserialized and loaded into the in-memory \textit{graph cache} to support operator-level processing. To maintain the correspondence between topological structures and their associated vertex and edge records in the record storage, we further design a set of mapping structures within the topology storage. These mappings are initialized at system startup and made available to the execution engine for consistent and efficient graph data access. We list them as follows:
\begin{itemize}[left=0pt]
    \item \textit{nidMap}: given the key of a vertex record, retrieve the corresponding node ID ($nid$) in the adjacency graph. \\
    input = ($oid$, $vid$); output = $nid$.
    
    \item \textit{vertexMap}: given an $nid$ in the adjacency graph, retrieve the location of the corresponding vertex in the record storage. \\
    input = $nid$; output = $(oid,  tid)$.

    \item \textit{edgeMap}: given the source and target node IDs, retrieve the location of the corresponding edge in the record storage. \\
    input = ($nid_s$, $nid_t$); output = $(oid,  tid)$.
\end{itemize}
In the mapping structures, parameter \textit{tid} denotes the tuple identifier automatically assigned by \mmdb to each record. Retrieving records via \textit{tid} can avoid table scan and guarantee constant time complexity $O(1)$.

\myparagraph{Inter-buffer for Matrix Storage}
To address the inefficiencies of tuple-at-a-time processing in \ta under the volcano model, we introduce an in-memory inter-buffer that materializes intermediate data for batched \ta processing. The inter-buffer is used as an internal structure for analytical operator invocation and is not exposed in standard query processing. This buffer organizes data into a matrix-oriented layout, which is more suitable for executing complex analytical computations. Specifically, \mmdb supports two strategies for matrix construction: 1) \textit{local access}, which extracts numerical attributes from relational tables and directly assembles them into a matrix, and 2) \textit{random access}, which constructs matrices by aggregating multi-valued attributes from qualifying records based on filtering conditions. The operators responsible for matrix generation and \ta are discussed in Section~\ref{sec: operators}.

\subsection{Data Access Methods}
\label{sec: Data Access Methods}

\myparagraph{Record Access Methods}
In \mmdb, all records retrieved from the record storage are loaded into the in-memory row buffer pool. Records that satisfy a given \textit{predicate} are then passed to the query processing engine, where a predicate can be defined as follows:

\begin{definition}[\textbf{{Predicate}}] \label{def: predicate}
A predicate is defined as a function $\mathcal{F}: a_1 \times \dots \times a_m \to \{ \texttt{True}, \texttt{False} \}$ applied to a record (or a pair of records in the case of a join) with $m$ attributes.
\end{definition}
To access records, \mmdb employs two record access methods (RecordAMs): a scan-based method and a $tid$-based method. As these techniques are well established in existing relational database systems \cite{PostgreSQL,MySQL}, we omit their detailed description. Instead, we emphasize that during graph processing, \mmdb favors $tid$-based access whenever possible to reduce the overhead of large-scale scans. Moreover, the unified RecordAMs eliminate model-specific access paths and provide a consistent interface for reading and writing multi-model data.




\myparagraph{Graph Access Methods}
The usage of \ag in \mmdb fundamentally differs from that of predefined join indexes or adjacency lists in existing GNSs, which primarily serve pure topology-based graph traversal. Instead, \mmdb retrieves nodes from adjacency graphs through newly designed graph access methods (GraphAMs) that enable hybrid traversal over both record attributes stored in the record storage and graph topology. GraphAMs retrieve candidate nodes by jointly navigating the adjacency graph and evaluating attribute predicates on associated records. The retrieved results are then passed to graph operators, supporting both topology-only queries (e.g., shortest-path search) and hybrid queries that combine topological constraints with property evaluation (e.g., pattern matching). Importantly, GraphAMs are encapsulated as a novel \textit{hybrid traversal operator}, which forms a foundational building block of \mmdb’s operator framework. This design integrates GraphAMs into the unified operator library (Section~\ref{sec: operators}), enabling \mmdb to support advanced \ti optimizations.

\subsection{Data Updates and Consistency Control}
Although \mmdb is primarily tailored for \task, we also provide a basic consistency control mechanism to support graph update operations. In this work, we focus on the following three types of updates.

\myparagraph{Update}
In a property graph $G$, property updates modify only the attribute fields of vertices and edges, i.e., the \texttt{props} associated with $V$ and $E$, without altering the graph topology. Following the conventional update semantics of relational database management systems (RDBMSs), \mmdb applies such updates directly to the record storage through RecordAMs, while leaving the topology storage unchanged.

\myparagraph{Insertion}
Batch insertion of vertices and edges introduces two main challenges: preserving graph topological consistency and maintaining synchronization between the record storage and the topology storage. \mmdb addresses these challenges using a staged insertion protocol. Specifically, records are first inserted into the record storage via RecordAMs. For newly inserted vertices, \mmdb then allocates corresponding topology nodes with fresh $nid$s in the topology storage, and incrementally updates the adjacency graph as well as the record-topology mappers described in Section~\ref{sec: Data Storage Format}. Edge insertions are handled by updating the adjacency relationships between the corresponding source and target topology nodes. For vertex-only insertions without incident edges, \mmdb optimizes the insertion workflow by omitting adjacency updates, thereby reducing insertion overhead while preserving consistency across the dual storage layers.

\myparagraph{Deletion}
Deleting vertices or edges may invalidate adjacency relationships and compromise query correctness. To handle deletions consistently, \mmdb performs deletion through the record-topology mappers. Specifically, when a vertex or edge is deleted, \mmdb first removes the corresponding topology nodes or adjacency entries from the topology storage using the maintained mappers. It then deletes the associated mapper entries and records from the record storage. This procedure ensures that both the topology storage and the record storage remain logically consistent. By localizing deletions to the affected topology nodes and mapper entries, \mmdb avoids global recomputation of the adjacency graph while preserving correctness for subsequent \ti execution.

%% file: 3Definitions.tex
\section{SCALABLE OPERATOR LIBRARY}
\label{sec: operators}
In this section, we present the operator library built atop the storage engine. By invoking specialized operators for data access and query execution to support \task, \mmdb overcomes the functional and performance limitations of handling multi-model data, as well as \ti and \ta, within a single database.

\input{T2D}

\subsection{Hybrid Traversal Operator}\label{sec: Hybrid Traversal Operator}
\begin{figure}[t]
    \centering
    \includegraphics[width=0.48\textwidth]{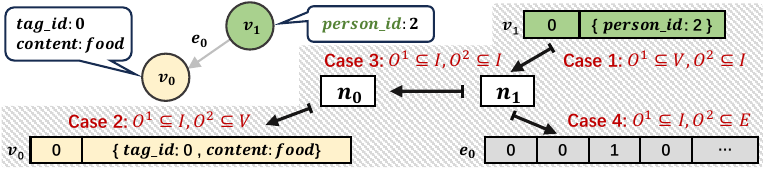}
    \caption{An example of hybrid traversal operations.}
    \label{fig: traversal}
\end{figure}

The design of topology- and attribute-aware graph operators brings two key advantages: 1) it enables efficient graph traversal while supporting predicate pushdown, allowing topological exploration and attribute filtering to be performed jointly during execution, and 2) attribute-aware traversal allows graph operators to directly interact with records from other data models, creating new opportunities for optimizations in \ti. To support such data interaction within the dual storage engine, we design a hybrid traversal operator (as GraphAM) that executes queries involving both topology and property access by linking the adjacency graph $\Omega = (N_s, N_t, \mathcal{I})$ with the corresponding vertex and edge records of graph $G$ in the record storage. 
The operator, denoted as $\mapsto$, is a binary operator that takes two operands $O^1$ and $O^2$ as inputs, and emits each valid pair $(r^1, r^2) \in O^1 \times O^2$ as outputs. The operand type takes one of the following four combinations:
\begin{equation} \label{eq: hybrid traversal}
O^1 \times O^2 \in \{ V \times I, I \times V, I \times I, I \times E \},
\end{equation}
where $V$ and $E$ denote the vertex and edge record sets, and $I$ denotes the full $nid$ set, i.e., $\mathcal{I}(N_s)$, in the \ag. We explain these four operand combinations in Figure~\ref{fig: traversal}:
\begin{itemize}[left=0pt]
    \item Case 1: traversal from vertex records to $nid$s;
    \item Case 2: traversal from $nid$s to vertex records;
    \item Case 3: traversal from source $nid$s to target $nid$s;
    \item Case 4: traversal from source $nid$s to edge records.
\end{itemize}

Algorithm~\ref{alg: T2D} illustrates the execution procedure of the hybrid traversal operator. Each candidate result pair $(r^1, r^2)$ is maintained in a queue structure (Line~\ref{alg1: line1}). The conditional branches in Line~\ref{alg1: line2}–\ref{alg1: line25} implement the four cases described in Eq.~\eqref{eq: hybrid traversal}. The operator follows the volcano iterator model \cite{graefe1994volcano}, which invokes an $emit()$ function to retrieve the next result pair from the queue (Lines~\ref{alg1: line32}-\ref{alg1: line33}). Although the membership test in Line~\ref{alg1: line17} (i.e., $nid_t \in O^2$) may appear expensive at first glance, it is fundamentally different from record-level joins in TBSs. Here, both $nid_t$ and $O^2$ are in-memory symbolic identifiers rather than records, and the check corresponds to a lightweight identifier membership test with no I/O access. Moreover, \mmdb applies additional lightweight optimizations to avoid unnecessary $O(|O^1| \cdot |O^2|)$ traversals in such cases, which will be introduced in Section~\ref{sec: pattern matching}. This operator unifies graph topology and property evaluation by integrating structural navigation with attribute filtering. This integration underpins \ti optimization and supports graph operators, such as shortest-path search. Due to space constraints, we focus on the pattern matching operator in the following section, which directly participates in \ti planning.

\subsection{Optimized Pattern Matching Operator} 
\label{sec: pattern matching}
The essence of pattern matching is extracting subsets of vertex and edge records that satisfy the topological and property constraints in a pattern $P$. Many MMDBs \cite{agensgraph,Hassan2018Extending,lin2016fast,luo2025relgo} aggressively push attribute predicates before the pure topology traversal (or join) phase. However, for predicates with low selectivity reduction (e.g., inequality predicates), such early predicate pushdown does not effectively reduce the number of intermediate results. Instead, it introduces additional scanning and evaluation overhead, which may outweigh the potential benefits.
To accelerate this process, we propose a novel topology- and attribute-aware pattern matching operator, which determines whether to push down attribute predicates during graph processing based on a cost-based decision. Built upon the hybrid traversal operator, it inherits efficient interaction between records and topology, thereby enabling greater potential for query optimization (see Section~\ref{sec: query optimization}).

\begin{figure}[t]
    \centering
    \includegraphics[width=0.48\textwidth]{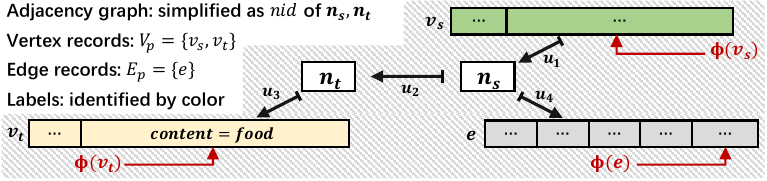}
    \caption{An example of a pattern.}
    \label{fig: pattern}
\end{figure}

\myparagraph{Definition}
Firstly, we define a graph \textit{pattern} as $P=(G_p, U, \Phi)$, where $G_p=(\Omega_p, V_p, E_p, \mathcal{L}_p)$ is the pattern graph, $U=[u_1,\dots,u_x]$ is an ordered sequence of $x$ hybrid traversal operations with each $u=(O^1 \mapsto O^2) \in U$ emitting exactly one result pair, and $\Phi:V_p \cup E_p \to \{\mathcal{F}_1,\dots,\mathcal{F}_y\}$ $(y=|V_p|+|E_p|)$ denotes a predicate assignment function that maps record-level predicates to vertices and edges.
For instance, Figure~\ref{fig: pattern} shows a simple vertex-edge-vertex pattern, with $U=[u_1, u_2, u_3, u_4]$, of the following expression, where the property evaluation predicates in the \texttt{WHERE} clause are represented by the predicate assignment function $\Phi(\cdot)$ (the red arrows in Figure~\ref{fig: pattern}):

{\small
\begin{verbatim}
MATCH (p:Persons)-[e:Interested in]->(t:Tags)
WHERE t.content = `food'
\end{verbatim}
}

\input{pattern_match}

\myparagraph{Topology-aware Algorithm Design}
We present the pseudocode of pattern matching operator $\mathcal{P}(G, P)$ in Algorithm~\ref{alg: pattern match}. The algorithm begins by constructing a candidate mapping function $\mathcal{M}(\cdot)$ (Lines~\ref{algo2: line3}--\ref{algo2: line8}), which maps each pattern vertex, pattern edge, and their corresponding $\textit{nid}$s to candidate records (or $\textit{nid}$s) in $G$. In practice, records stored in the same collection share identical labels. Therefore, Lines~\ref{algo2: line4} and~\ref{algo2: line8} reduce to selecting vertex or edge record sets with matching labels, without requiring additional scanning.
After initializing the candidate sets, the algorithm verifies the topological constraints by executing the ordered operation sequence $U$ defined in the pattern (Lines~\ref{algo2: line11}–\ref{algo2: line22}).
The traversal is initiated from each candidate record of the first operand (Line~\ref{algo2: line12}), and the system incrementally applies the hybrid traversal operators $(O_i^1 \mapsto O_i^2)$ in the order specified by $U$.
During execution, a stack $S$ is maintained to track currently valid partial paths (Line~\ref{algo2: line13}).
At each step, the current intermediate result is expanded by invoking the corresponding hybrid traversal operator, which retrieves valid topological neighbors via adjacency access and emits matching record pairs.
Once all operations in $U$ are successfully applied, the accumulated records along the path are materialized into the result sets $(V_m, E_m)$. The output $(V_m, E_m)$ forms a \textit{graph-relation}, which is materialized as an intermediate result collection in \mmdb and can be directly consumed by subsequent \ti operators.

\begin{figure}[t]
    \centering
    \includegraphics[width=0.48\textwidth]{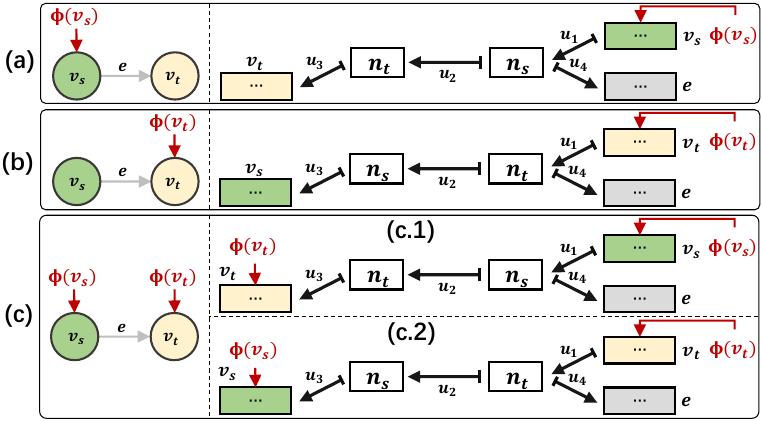}
    \caption{Optimization mechanisms for patterns with different predicate types. The left of the dashed line shows the types represented by the pattern, while the right shows \mmdb’s candidate or executed plans.}
    \label{fig: attribute}
\end{figure}

\myparagraph{Attribute-aware Optimization}
We design a systematic set of rule-based and cost-based predicate pushdown strategies to handle attribute predicates in graph queries, and summarize them in Figure~\ref{fig: attribute}. For clarity, we illustrate these strategies using a simplified vertex-edge-vertex topology and focus on predicates on the first and last vertices, which guide the planner’s choice between forward and reverse traversal.

Figure~\ref{fig: attribute}(a) and (b) illustrate patterns with a single attribute predicate on either the source or target vertex. For such cases, we adopt a rule-based strategy: \textit{push down the predicate and then initiate traversal from the predicate side}. Without this strategy, executing the pattern in Figure~\ref{fig: attribute}(b) in the original direction leads to inefficiencies: applying topology evaluation before the predicate retrieves invalid records and incurs unnecessary I/O, while applying the predicate first requires an $O(|O^1|\cdot|O^2|)$ membership test during traversal. By starting traversal from the predicate-filtered vertex $v_t$, all generated neighbors are guaranteed to be valid, and the membership test at the opposite end is guaranteed to always succeed and can therefore be safely omitted.

Figure~\ref{fig: attribute}(c) shows patterns where both the source and target vertices carry attribute predicates, yielding two candidate traversal directions, as shown in Figure~\ref{fig: attribute}(c.1) and (c.2).
We estimate the effective cardinalities after predicate filtering as $|\mathcal{M}(v_s)| \times \mathcal{S}_{\Phi(v_s)}$ and $|\mathcal{M}(v_t)| \times \mathcal{S}_{\Phi(v_t)}$, where $\mathcal{M}(\cdot)$ denotes the mapping function defined in Section~\ref{sec: Hybrid Traversal Operator} and $\mathcal{S}$ represents predicate selectivity.
The predicate with the smaller estimated cardinality is pushed down and chosen as the traversal starting point.
For predicates on the end vertex, we apply the following rules: \textit{equality predicates are always pushed down, while inequality predicates are deferred until after topology traversal}. For range predicates, we estimate and compare the costs of pushing down versus postponing the predicate, and select the lower-cost option in the execution plan. The cost model is discussed in Section~\ref{sec: cost model}. Note that once the predicates are pushed down, Lines~\ref{algo2: line4} and~\ref{algo2: line8} in Algorithm~\ref{alg: pattern match} are changed to $\mathcal{M}(v_p) \gets \{ v \in V | \mathcal{L}(v) = \mathcal{L}_p(v_p), \Phi(v_p)(v) = \texttt{True}\}$ and $\mathcal{M}(e_p) \gets \{ e \in E | \mathcal{L}(e) = \mathcal{L}_p(e_p), \Phi(e_p)(e) = \texttt{True}\}$.

\subsection{Cross-model Join Operator}
Current MMDBs \cite{duckpgq,ArangoDB2025,agensgraph,Hassan2018Extending} struggle with \ti, as they process different data models separately and lack explicit operators for correlating records across models during query execution. To bridge this gap, we introduce a \textit{cross-model join operator} that directly associates relational, document, and graph models.
A cross-model join $H^1 \widehat{\bowtie}_{\mathcal{F}} H^2$ links two data collections, $H^1, H^2 \in \{ R, D, G \}$, based on the given predicate $\mathcal{F}(\cdot)$. The operator adopts one of two execution strategies for join processing, depending on the data models involved: 1) the join between relational and document collections, and 2) the join between a graph and a relational or document collection.

Algorithm~\ref{alg: join} presents the pseudocode of the cross-model join procedure. In our design, the join between relational and document collections (Line~\ref{algo3: line2}–\ref{algo3: line3}) directly links record entities from these two distinct data models. The NF$^2$ constraint in \mmdb naturally accommodates the storage of intermediate join results. For the join between a graph and a relational or document collection (Line~\ref{algo3: line4}–\ref{algo3: line12}), the output remains in the form of a graph, but with modified vertex or edge record sets resulting from the join. We do not design join operations between two graphs, although the intuitive result of such a join would be a larger graph. In \mmdb, this functionality can already be expressed through two match operations, which effectively leverage the existing operator framework.

\input{join}

\subsection{Parallel Analytical Operators}\label{sec: Analysis Operators}
To address the functional and performance limitations of tuple-at-a-time processing for \ta, we introduce a suite of parallel analytical operators (summarized in Table~\ref{tab: array operators}) that reorganize \ta execution into a batch-processing pipeline.

\myparagraph{Matrix Generation}
As described in Section~\ref{sec: Data Storage Format}, we design two strategies for matrix generation: local access and random access. For local access, which extracts certain fields from relational tables (or intermediate views) without predicates and transforms them into a matrix, we implement a \texttt{REL2MATRIX} operator that bypasses tuple-at-a-time scans by directly reading columnar data and stores the output in the {inter-buffer}. For random access, we implement it by invoking the scan-based RecordAM defined in Section~\ref{sec: Data Access Methods}.

\myparagraph{Linear Algebra}
In \mmdb, analytical operators are built atop a block-based parallel algebraic architecture, which decomposes large matrices into independently executable sub-blocks (tiles). This design addresses a long-standing challenge of \ta: limited parallel granularity under tuple-at-a-time execution models. Each block is dynamically scheduled across worker threads within a shared-memory framework for \ta such as scalable matrix multiplication, cosine similarity, and logistic regression. For example, given two matrices $\textbf{X}$ and $\textbf{Y}$, the multiplication $\textbf{Z} = \textbf{X} \cdot \textbf{Y}$ is performed by computing block-wise partial products $\textbf{Z}_{ij} = \sum_k \textbf{X}_{ik} \cdot \textbf{Y}_{kj}$, where each $(i,j)$ block is processed independently. Similarly, cosine similarity is computed via distributed inner products and normalization across row vectors, while logistic regression involves iterative gradient computation aggregating contributions from each partition in parallel. All intermediate results are exchanged via \mmdb’s in-memory inter-buffer, minimizing disk I/O and enabling synchronization.

Note that the operator library is designed to be extensible, allowing developers to easily implement specific physical operators for \ta and seamlessly integrate them into \mmdb.

\input{array_operators}

%% file: T2D.tex
\begin{algorithm}[t]\small
\SetArgSty{textnormal}
\SetAlgoLined
	\caption{Hybrid Traversal $O^1 \mapsto O^2$}
        \KwIn{$O^1$: the first operand, $O^2$: the second operand}
	\label{alg: T2D}

        Initialize $res$ as an empty queue\;    \label{alg1: line1}

        \If{$O^1 \subseteq V$ and $O^2 \subseteq I$}{ \label{alg1: line2}
            \ForEach{$v \in O^1$}{
                Get the $(oid, tid)$ of $v$\;
                $nid \gets$ $\textit{nidMap}(oid, tid)$\;
                Push back $(v, nid)$ to $res$\;
            }
        }
        \If{$O^1 \subseteq I$ and $O^2 \subseteq V$}{
            \ForEach{$nid \in O^1$}{
                $(oid, tid) \leftarrow \textit{vertexMap}(nid)$\;
                Get vertex record $v$ via $tid$-based RecordAM\;
                Push back $(nid, v)$ to $res$\;
            }
        }
        \If{$O^1 \subseteq I$ and $O^2 \subseteq I$}{
            \ForEach{$nid_s \in O^1$}{
                Find $n_s \in N_s$ with $\mathcal{I}(n_s) = nid_s$\;
                \While{$n_s.next \neq \texttt{Null}$}{
                    $n_t \gets n_s.next$, $nid_t \gets \mathcal{I}(n_s.next)$\;
                    Push back $(nid_s, nid_t)$ to $res$ \textbf{if} $nid_t \in O^2$\; \label{alg1: line17}
                    $n_s \gets n_t$\;
                }
            }
        }
        \If{$O^1 \subseteq I$ and $O^2 \subseteq E$}{
            \ForEach{$nid_s \in O^1$}{
                \While{$(nid_s \mapsto \mathcal{I}(N_s)).emit() \neq \texttt{Null}$}{
                    $res \gets (nid_s \mapsto \mathcal{I}(N_s)).emit()$\; 
                    $(oid, tid) \gets \textit{edgeMap}(res)$\;
                    Get record $e$ via $tid$-based RecordAM\;
                    Push back $(nid, e)$ to $res$ \textbf{if} $e \in O^2$\; \label{alg1: line25}
                }
            }
        }

    \If{the $emit()$ function is involved}{     \label{alg1: line32}
        Pop and emit the front element form $res$\;     \label{alg1: line33}
    }

\end{algorithm}

%% file: pattern_match.tex
\begin{algorithm}[t]\small
\SetArgSty{textnormal}
\SetAlgoLined
	\caption{Pattern Matching $\mathcal{P}(G,P)$}
    \KwIn{$G = (\Omega, V, E, \mathcal{L})$: a graph, $P = (G_p, U, \Phi)$: a pattern}
	\KwOut{$(V_m, E_m)$: matched vertex and edge records sets}
	\label{alg: pattern match}

    Initialize $V_m$ and $E_m$ as empty sets\;  \label{algo2: line1}

    Initialize a mapping function $\mathcal{M}$\;    \label{algo2: line2}

    \ForEach{pattern vertex $v_p \in V_p$ and its $nid_p$}{\label{algo2: line3}
        $\mathcal{M}(v_p) \gets \{ v \in V \ | \ \mathcal{L}(v) = \mathcal{L}_p(v_p)\}$; \label{algo2: line4}

        $\mathcal{M}(\textit{nid}_p) \gets \{ \textit{nidMap}(v) \ | \ v \in \mathcal{M}(v_p) \}$\;
    }
    \ForEach{pattern edge $e_p \in E_p$}{
        $\mathcal{M}(e_p) \gets \{ e \in E \ | \ \mathcal{L}(e) = \mathcal{L}_p(e_p)\}$;\label{algo2: line8}
    }

    Get the first operation $u_1 = (O_1^1 \mapsto O_1^2)$ from $U$\; \label{algo2: line11}

    \ForEach{$v_0 \in \mathcal{M}(O_1^1)$}{ \label{algo2: line12}
        Initialize an empty stack $S \gets [(v_0, 1, [v_0])]$\;\label{algo2: line13}
    
        \While{$S$ is not empty}{
            Pop $(current, i, path)$ from $S$\;
    
            \If{$i > |U|$}{
                Add records from $path$ to $V_m$ and $E_m$\;
                \textbf{continue}\;
            }
            Get the $i$-th operation $u_i = (O_i^1 \mapsto O_i^2) \in U$\;

            \While{$res \gets (current \mapsto \mathcal{M}(O_i^2)).emit()$}{
                $(r^1, r^2) \gets res$\;
                Push $(r^2, i+1, path + [r^2])$ to $S$\;    \label{algo2: line22}
            }
        }
    }
    
    \KwRet $(V_m, E_m)$ \;

\end{algorithm}

%% file: join.tex
\begin{algorithm}[t]\small
\SetArgSty{textnormal}
\SetAlgoLined
\caption{Cross-model Join $H^1 \widehat{\bowtie}_{\mathcal{F}} H^2$}
\KwIn{$H^1, H^2$: two data collections, $\mathcal{F}$: a predicate}
\KwOut{$H$: a linked data collection}
\label{alg: join}

Initialize $H$ as an empty data collection\;  

\If{$H^1, H^2 \in \{ R, D \}$}{     \label{algo3: line2}
    $H \gets \{(h^1, h^2) \in H^1 \times H^2 \ | \ \mathcal{F}(h^1, h^2) = \texttt{True} \}$\;     \label{algo3: line3}
}

\If{$H^1\in \{ R, D \}$ and $H^2 = G$}{     \label{algo3: line4}
    \If{$\mathcal{F}$ is a predicate on $H^1$ and vertices of $G$}{
        Get the vertex record set $V$ from $G$\;
        $V \gets \{(h^1, v) \in H^1 \times V  \ | \ \mathcal{F}(h^1, v) = \texttt{True} \}$\;
        Update $G$ with $V$, and $H \gets G$\;
    }
    \If{$\mathcal{F}$ is a predicate on $H^1$ and edges of $G$}{
        Get the edge record set $E$ from $G$\;
        $E \gets \{(h^1, e) \in H^1 \times E \ | \ \mathcal{F}(h^1, e) = \texttt{True} \}$\;
        Update $G$ with $E$, and $H \gets G$\;  \label{algo3: line12}
    }
}

\KwRet $H$\;

\end{algorithm}

%% file: array_operators.tex
\begin{table}[t] \small
\setlength{\tabcolsep}{6pt}
\ra{1.0}
\centering
\caption{Analysis operators designed in \mmdb.}
\label{tab: array operators}
\begin{tabular}{cc}
\toprule
\textbf{Operators}           & \textbf{Description}                  \\ \midrule
\texttt{REL2MATRIX}      & Transform a relational table into a matrix    \\
\texttt{MULTIPLY}            & Perform matrix multiplication                  \\
\texttt{SIMILARITY}   & Compute cosine similarity                     \\
\texttt{REGRESSION} & Train a logistic regression model            \\

\bottomrule
\end{tabular}
\end{table}

%% file: 5Optimizations.tex
\section{UNIFIED QUERY PROCESSING ENGINE}
\label{sec: execution engine}

\subsection{\ti Plan Generation} \label{sec: cross-model Query Plan}

We achieve plan generation and physical execution of multi-model data within a unified processing engine by introducing novel physical operators. In traditional RDBMSs, a \texttt{Select}-\texttt{From}-\texttt{Where} query is first transformed by the parser into a query tree, and then the optimizer generates a physical execution plan. To support SFMW-based \ti, we implement and integrate the pattern matching operator (Section~\ref{sec: pattern matching}) into \mmdb as a new physical operator, thereby realizing the match operation in Eq.~\eqref{eq: sfmw}. 

We further introduce a new \texttt{GRAPH\_SCAN} operator into \mmdb to support the graph projection operation $\widehat{\pi}_{A'}$ in Eq.~\eqref{eq: sfmw}. The \texttt{GRAPH\_SCAN} operator encapsulates the scan-based RecordAM from RDBMSs, and is specialized for scanning and filtering graph vertex, edge records, and graph-relations produced by graph queries. Since both its input and output are organized as relational structures, \texttt{GRAPH\_SCAN} enables seamless integration of graph query plans with those over relational and document data, thereby facilitating unified multi-model query optimization in Section~\ref{sec: query optimization}.

\subsection{\ti Optimization}
\label{sec: query optimization}
In this section, we present the first optimization framework for \ti, which contains four novel mechanisms: 1) graph predicate pushdown, 2) join pushdown, 3) \ti rewriting, and 4) query-aware traversal pruning.

\myparagraph{Graph Predicate Pushdown}
\mmdb employs two novel graph predicate pushdown mechanisms that selectively apply predicates at the lowest levels of the query tree within the graph, thereby reducing the search space at an early stage and enabling \textit{predicate-level interaction} across multi-model data. The first mechanism pushes predicates into the match operation, as described in Section~\ref{sec: pattern matching}, where predicates are pushed down based on the designed rule-based and cost-based mechanisms.
The second mechanism is triggered when processing \ti that exhibit the following structure:
\begin{equation} \label{eq: predicate pushdown}
T_{\ti} = \pi_A(\sigma_\Psi(H_1 \widehat{\bowtie}_{\mathcal{F}_1} H_2 \widehat{\bowtie}_{\mathcal{F}_2} (\widehat{\pi}_{A'} \mathcal{P}(H_3, P)))),
\end{equation}
where $H_1$ and $H_2$ are relational or document collections, and $H_3$ denotes a graph. A predicate $\mathcal{F} \in \Psi$ is pushed down to $H_3$ if it falls into one of the following two cases:
\begin{itemize}[left=0pt]
    \item $\mathcal{F}$ is a predicate on $H_3$ that is not included in $P$, in which case we push down $\mathcal{F}$ into $H_3$ and update $\Psi \gets \Psi \setminus \{\mathcal{F}\}$;
    \item $\mathcal{F}$ is originally defined on $H_1$ or $H_2$, but $H_3$ shares the attribute referenced by $\mathcal{F}$, in which case we replicate and apply $\mathcal{F}$ to $H_3$.
\end{itemize}
This mechanism targets queries in which constraints are not incorporated into the graph pattern itself but are instead enforced subsequently through selection operators applied to the graph-relational results of pattern matching.

\myparagraph{Join Pushdown}
Join order optimization has been extensively studied in RDBMSs, yet no corresponding optimization rules have been developed for \ti due to the independent execution across models. To bridge this gap, we introduce a novel cost-based join pushdown rule that enables effective join order optimization in \ti evaluation. Specifically, given the \ti in Eq.~\eqref{eq: predicate pushdown}, \mmdb can generate a semantically equivalent candidate plan as follows:
\begin{equation} \label{eq: join pushdown}
T_{\ti} = \pi_A(\sigma_\Psi(H_1 \widehat{\bowtie}_{\mathcal{F}_1} (\widehat{\pi}_{A'} \mathcal{P}(H_2 \widehat{\bowtie}_{\mathcal{F}_2} H_3, P)))),
\end{equation}
which transforms the join between $H_2$ and a graph-relation into a join between $H_2$ and the graph $H_3$, leveraging the cross-model join operator to support seamless \textit{record-level interaction} across multi-model data. The final execution plan is then selected using a cost-based method, with the cost model presented in Section~\ref{sec: cost model}. Moreover, \mmdb can also push down $H_1$ into the match operation, generating another semantically equivalent candidate:
\begin{equation} \label{eq: join pushdown2}
T_{\ti} = \pi_A(\sigma_\Psi (\widehat{\pi}_{A'} \mathcal{P}(H_1 \widehat{\bowtie}_{\mathcal{F}_1} H_2 \widehat{\bowtie}_{\mathcal{F}_2} H_3, P))),
\end{equation}
where all cross-model join operations are executed before matching.

\myparagraph{\ti Rewriting}
With the integration of graph operators and the generation of \ti plans, we further contribute by designing a set of novel \ti rewriting rules, containing \textit{match trimming} and \textit{projection trimming}. Specifically, the match trimming rule enables \mmdb to rewrite match operations into simplified record-scan plans under specific conditions, thereby reducing execution overhead. In particular, rewriting applies when: 1) the pattern involves no topological constraints and requires only property evaluation on vertices or edges, or 2) the pattern contains only vertex-edge-vertex structure, with property evaluation applied solely to the edges. The following examples illustrate these two cases:

{\small
\begin{verbatim}
1) SELECT v2 FROM Interested_in
   MATCH (v2: Tags)
   WHERE v2.content = `food'
2) SELECT e FROM Interested_in
   MATCH (v1:Persons)-[e:Interested in]->(v2:Tags)
   WHERE e.weight > 0.5
\end{verbatim}}

Furthermore, the projection trimming rule eliminates a column subset $A'' \subseteq A'$ from a graph projection $\widehat{\pi}_{A'}$ if and only if the attributes in $A''$ are not referenced by any subsequent projection operations or predicates in \ti formulated as Eq.~\eqref{eq: sfmw}. By removing such redundant attributes, this rule reduces unnecessary memory consumption during \ti execution and simplifies the input for the subsequent query-aware traversal pruning mechanism.

\myparagraph{Query-aware Traversal Pruning}
In a graph pattern matching query with graph projection $\widehat{\pi}_{A'} \mathcal{P}(G, P)$, predicates in $P$ typically concern only a subset of vertices and edges, as not all vertices and edges in the pattern topology participate in property evaluation or projection. To reduce the I/O overhead of retrieving such irrelevant records, we introduce a query-aware traversal pruning mechanism. Specifically, match operations are realized through an ordered sequence $U$ of hybrid traversal operations (see Section~\ref{sec: pattern matching}). We prune a hybrid traversal $u = (O^1 \mapsto O^2) \in U$ if it satisfies the following conditions: $O^1 \times O^2 \in \{I \times V, I \times E\}$, $O^2$ is not included in $\widehat{\pi}_{A'}$, and $\Phi(O^2) = \texttt{Null}$. This corresponds to hybrid traversal operations like $u_3$ and $u_4$ in Figure~\ref{fig: pattern}. As a result, the pruned pattern matching can largely be executed directly over \ag, thereby reducing the I/O cost of fetching concrete records.

\subsection{Cost Model}
\label{sec: cost model}
In this section, we present the cost model designed for graph operators and \ti planning, where cost is measured in terms of disk I/O and CPU computation, following conventional disk-based databases \cite{balmin2006cost,graefe1993volcano}. 
Assume that each disk access incurs a cost of $Cost_{{I/O}}$ and each function call or predicate evaluation incurs a CPU cost of $Cost_{cpu}$.

\myparagraph{Cost for Hybrid Traversal}
The cost of a hybrid traversal operation $O^1 \mapsto O^2$ can take four forms, corresponding to the four combinations of its operands. We discuss them separately:
\begin{itemize}[left=0pt]
    \item For Case 1, \mmdb retrieves the $nid$s of given vertex records in $O^1$ via one-to-one mappers, incurring $|O^1|$ function calls. We set $Cost_{\mapsto} = |O^1| \times {Cost}_{cpu}$.
    \item For Case 2, we first locate the records and then retrieve them based on the given $nid$s. In this case, the cost is estimated as $|O^1| \times ({Cost}_{cpu} + {Cost}_{I/O})$.
    \item Case 3 retrieves adjacent node $nid$s. The cost is proportional to the average out-degree of the graph $G=(\Omega, V, E, \mathcal{L})$ and is estimated as $\frac{|O^1| \times |E|}{|V|} \times {Cost}_{cpu}$.
    \item Case 4 first retrieves adjacent node $nid$s and then accesses the corresponding edge records. The cost is estimated as $Cost_{\mapsto} = \frac{|O^1| \times |E|}{|V|} \times (2{Cost}_{cpu} + {Cost}_{I/O})$
\end{itemize}

\myparagraph{Cost for Pattern Matching}
As discussed in Section~\ref{sec: pattern matching}, a pattern matching operation (Algorithm~\ref{alg: pattern match}) evaluates the costs of different execution strategies to decide whether attribute predicates should be pushed down. We now describe the corresponding cost model.

Let $|\Phi(V_p)|$ and $|\Phi(E_p)|$ denote the numbers of effective predicates on pattern vertices and edges, respectively, while $\alpha$ and $\beta$ denote the numbers of predicates pushed down to Lines~\ref{algo2: line4} and~\ref{algo2: line8} in Algorithm~\ref{alg: pattern match}, respectively. After executing Algorithm~\ref{alg: pattern match}, the remaining predicates are applied to the output graph-relation $(V_m, E_m)$. Their execution cost can be formulated as:
\begin{equation}
{Cost}_{\mathcal{P}} = Cost_{algo2} + Cost_{prop},
\end{equation}
where $Cost_{algo2}$ denotes the execution cost of Algorithm~\ref{alg: pattern match}, and $Cost_{prop}$ denotes the cost of the remaining property evaluation. 

Specifically, for each predicate that is pushed down to Lines~\ref{algo2: line4} and~\ref{algo2: line8} in Algorithm~\ref{alg: pattern match}, \mmdb incurs a cost of $(\alpha \cdot |V| + \beta \cdot |E|) \times (Cost_{I/O} + Cost_{cpu})$ to complete attribute evaluation. After that, a sequence of hybrid traversal operators is invoked for the source vertex set. Assuming an average of $\lambda$ hybrid traversal operations executed per record, the total cost of Algorithm~\ref{alg: pattern match} is:
\begin{equation}
\begin{aligned}
Cost_{algo2} =& (\alpha |V| + \beta |E|) 
                 \times (Cost_{I/O} + Cost_{cpu}) \\
               &+ \lambda \times Cost_{\mapsto}.
\end{aligned}
\end{equation}

We denote the number of rows in the resulting graph-relation $(V_m, E_m)$ as $|\mathcal{P}(G,P)|$. Under the attribute independence assumption \cite{leis2015job}, the cost of evaluating the remaining property predicates is:
\begin{equation}
Cost_{prop} = |\mathcal{P}(G,P)| \times Cost_{cpu}.
\end{equation}

\myparagraph{Cost for Cross-model Join}
Taking the nested loop join \cite{shin1994new} as an example, where the left and right input sizes are $N_L$ and $N_R$, respectively, the cost of a cross-model join is defined as follows:
\begin{equation}    \label{eq: join cost}
{Cost}_{\widehat{\bowtie}} = N_L \cdot N_R \cdot Cost_{cpu}.
\end{equation}
Eq.~\eqref{eq: join cost} assumes that both operand collections reside in the buffer pool and only captures the cost of linking all entities from both inputs. For disk-based scenarios, Eq.~\eqref{eq: join cost 1} presents the cost when the buffer pool can fully accommodate both operands, while Eq.~\eqref{eq: join cost 2} covers the case where only the left operand fits. Here, $b$ indicates the block capacity in terms of records.
\begin{gather}
{Cost}_{\widehat{\bowtie}} = \left(\tfrac{N_L}{b}+\tfrac{N_R}{b}\right) \cdot Cost_{I/O} + N_L \cdot N_R \cdot Cost_{cpu}, \label{eq: join cost 1} \\ 
{Cost}_{\widehat{\bowtie}} = \left(\tfrac{N_L}{b} + N_L \tfrac{N_R}{b}\right) \cdot Cost_{I/O} + N_L \cdot N_R \cdot Cost_{cpu}. \label{eq: join cost 2}
\end{gather}
Note that a join between a graph and another model performs entity linking between a vertex or edge record collection and the given data collection, and its cost can also be expressed using Eq.~\eqref{eq: join cost}-\eqref{eq: join cost 2}.

\subsection{\ta Pipeline}
\label{sec: analytical pipeline}
To support expressive \ta, \mmdb integrates a dedicated \ta pipeline that provides a unified framework for transforming \ti results into computation-ready analytical flows. Upon receiving a user-issued analytical task, \mmdb generates a structured pipeline plan that includes: 1) a logical ordering of operators, 2) transformation of intermediate results into matrices stored in the inter-buffer, and 3) parallel execution of \ta over inter-buffer contents.

\myparagraph{Operator Invocation Planning}
Constructing \ta pipelines is nontrivial, as analytical expressions depend on heterogeneous and interrelated outputs produced by \ti. To address this issue, \mmdb employs an operator invocation planner that analyzes fine-grained data dependencies exposed by the \ti parser and automatically derives an invocation plan that respects user-specified analytical semantics. This design ensures correct data flow across operators while enabling parallel, dependency-aware scheduling over multi-model inputs.

\myparagraph{Matrix Generation}
Once the logical order of analytical operator invocations is established, the engine inserts appropriate matrix generation operators to convert intermediate results into structured matrix representations. Two strategies are employed: local access, which directly maps numeric columns from base tables into matrices, and random access, which aggregates multi-valued attributes from filtered records. Both strategies materialize results into the in-memory inter-buffer, serving as a unified \ta workspace that bridges query outputs and computational operators.

\myparagraph{\ta Execution}
\mmdb schedules the execution of analytical operators, such as cosine similarity, over inter-buffer data using a block-based parallel workflow. By decomposing large matrices into independently processed tiles, \mmdb achieves scalable execution across multiple worker threads. The inter-buffer facilitates both operator input and result sharing, minimizing I/O and avoiding redundant computation. Intermediate results in the inter-buffer are reused across analytical tasks via structural matching of \ti plans, allowing semantically equivalent \task to share materialized outputs without re-execution. This modular architecture fundamentally replaces volcano-style execution with a parallel and scalable analytical framework, enabling efficient end-to-end \ta entirely within the database.

%% file: 6Experiments.tex
\section{EXPERIMENTS}
\label{sec:exp}
In this section, we evaluate the superiority of \mmdb. We first describe the experimental setup (Section~\ref{sec: setup}), followed by ablation studies that analyze the impact of individual optimization components (Section~\ref{sec: Ablation Studies}). We then present a comprehensive comparison between \mmdb and SOTA MMDBs on graph queries, \ti, and \ta (Section~\ref{sec: response time}-\ref{sec: Single-model workloads}).

\subsection{Experimental Setup} \label{sec: setup}
\myparagraph{Implementations}
We implemented \mmdb on top of the open-source RDBMS openGauss \cite{li2021opengauss}. All operators introduced in Section~\ref{sec: operators} are realized as physical operators and fully integrated into the system. The evaluation was performed on a high-end server featuring dual Intel Xeon Gold 5320 CPUs (2.20GHz, 52 cores/104 threads), 512GB of DDR4 memory, and a 3.7TB Samsung SSD 870. The system environment was based on CentOS Linux release 8.5.2111.

\begin{figure*}[t]
    \centering
    \includegraphics[width=\textwidth]{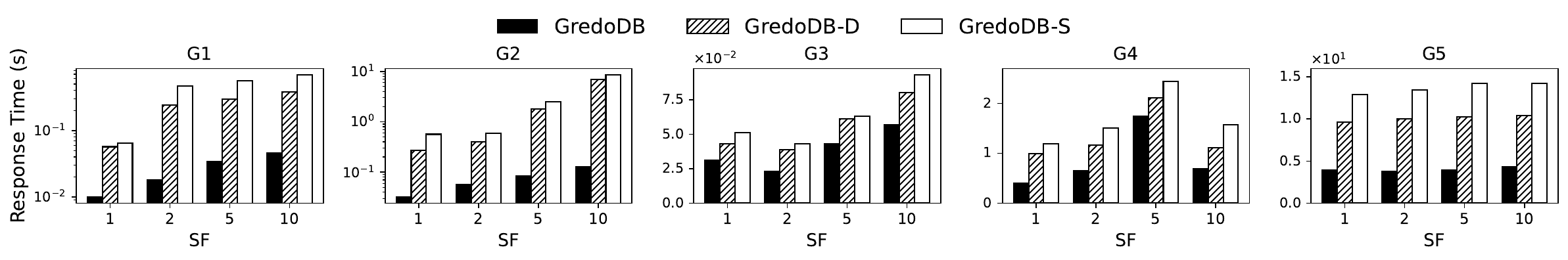}
    \caption{Response times (seconds) of different \mmdb variants on graph processing.}
    \label{fig: ablation graph}
\end{figure*}

\begin{figure*}[t]
    \centering
    \includegraphics[width=\textwidth]{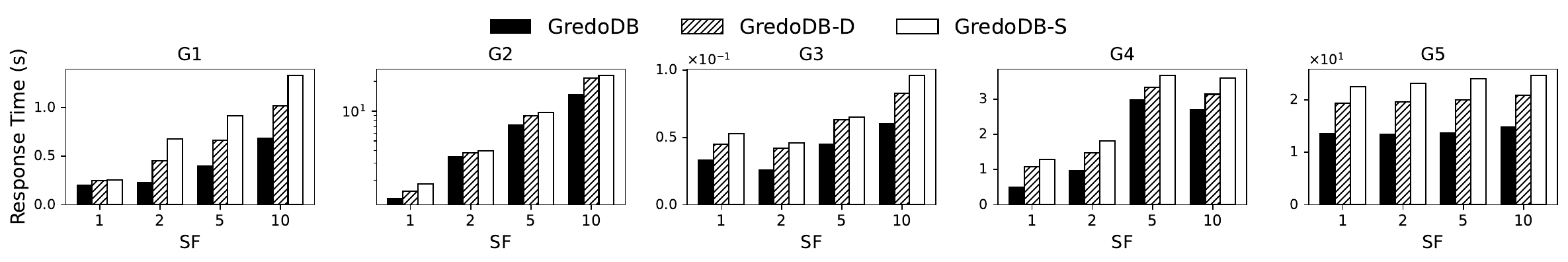}
    \caption{Response times (seconds) of different \mmdb variants on \ti.}
    \label{fig: ablation gcq}
\end{figure*}

\input{model}

\myparagraph{Benchmarks}
We adopt \bench \cite{kim2022m2bench}, the only multi-model benchmark tailored for \task, which consists of 17 queries (t0-t16) across three real-world scenarios. Note that the array model introduced in \cite{kim2022m2bench} is treated as an analytical workload in our evaluation, as \bench organizes the results of multi-model queries into array-like data structures and applies them to downstream analytical tasks. 
Table~\ref{tab: model} summarizes the workloads involved in different query tasks. For ease of reference in subsequent discussions, we assign symbolic aliases to the queries (e.g., R1-R7 correspond to queries involving the relational model, listed in order). 
Among the analytical queries defined in \bench, A1-A3 correspond to the \ta workloads that we focus on in this paper, while A4-A6 represent traditional aggregation-centric analytical queries commonly studied in the database literature.
To evaluate scalability and performance under varying data volumes, we conduct experiments using four scale factors (SF = 1, 2, 5, and 10). 

\myparagraph{Comparisons}
Based on our discussion in Section~\ref{sec: related work}, we select one representative system from each category for comparison:
\begin{itemize}[left=0pt]
    \item \poly: We build the \poly system, a representative MES following the design principles of \cite{kim2022m2bench}, relying on \texttt{MySQL} \cite{MySQL}, \texttt{Neo4j} \cite{neo4j}, \texttt{MongoDB} \cite{MongoDB2025MongoDB}, and \texttt{SciDB} \cite{Stonebraker2011scidb}, each of which natively supports a single model or analytical workload.
    \item \agens v2.1.3 \cite{agensgraph}: A representative TBS, natively supports relational and graph storage, with JSONB extension enabling document storage and join-index-based methods to accelerate graph queries. 
    \item \arango v3.12.4 \cite{ArangoDB2025}: A GNS that adopts documents as the first model, accelerates graph queries by building hash indexes on the \texttt{\_from} and \texttt{\_to} fields of edge collections to preserve the graph topology.

    \item \duck v1.3.2 \cite{duckdb}: a columnar OLAP database, with \duckpgq \cite{duckpgq} providing support for multi-join-based graph queries. Considering its parallel execution capabilities, we conduct experiments by restricting the number of threads to 1, 4, and 16, respectively.
\end{itemize}
All systems are configured with a shared buffer size of 30 GB and a work memory of 60 GB. We set the maximum query execution time to 5 hours, since most queries in \bench complete in far less time.

\subsection{Ablation Studies}
\label{sec: Ablation Studies}
We conduct a set of ablation studies to assess the effectiveness of key design components in \mmdb, including: 1) the dual storage engine and the graph operators for accelerating graph queries, 2) \ti-specific optimization techniques for improving \ti performance, and 3) the operator-level parallel execution architecture for reducing the execution time of \ta. 
Specifically, we evaluate the following three system variants:
\begin{itemize}[left=0pt]
    \item \mmdb, a dual-engine system, which uses \ag as input to specialized graph query operators and enables \ti optimization mechanisms.
    \item \mmdbd, a dual-engine system without the proposed optimization mechanisms, where graph queries are processed in a purely topology-driven manner (i.e., DFS or BFS), similar to \grfusion \cite{Hassan2018Extending}.
    \item \mmdbs, a single-engine system without the developed operators and optimization mechanisms, which executes \ti in the relational engine using primary-key-indexed joins.
\end{itemize}
Since the designed operators are tightly coupled with the optimizations proposed in Section~\ref{sec: query optimization}, we omit the ablation study separating operators from optimization techniques. For analytical workloads, both \mmdbs and \mmdbd rely on volcano model, resulting in the same response times in the following evaluations.

\begin{figure}[t]
    \centering
    \includegraphics[width=0.48\textwidth]{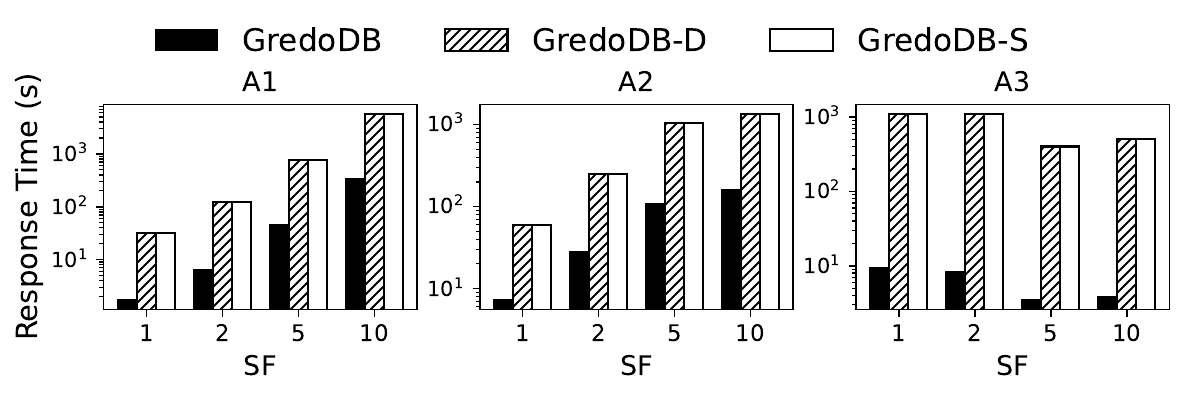}
    \caption{Response times (seconds) of different \mmdb variants on \ta workloads (A1-A3).}

    \label{fig: ablation analysis}
\end{figure}

\myparagraph{Optimization Effectiveness for Graph Processing and \ti}
We evaluate the effectiveness of graph processing and \ti execution in \mmdb, \mmdbd, and \mmdbs using G1-G5, which primarily focus on graph pattern matching. 
To isolate the impact of graph processing, we measure the execution time of graph-processing sub-plans as a proxy for graph processing efficiency, and report these results in Figure~\ref{fig: ablation graph}. The end-to-end response time of G1-G5 is recorded as overall \ti performance, as shown in Figure~\ref{fig: ablation gcq}. 
\ti tasks G6-G8 mainly evaluate shortest-path search, which is not supported by \mmdbd and \mmdbs, and are therefore excluded from this comparison.

\finding
(1) Both the dual-engine design and the architecture with specialized operators and optimizations accelerate graph and \ti processing, enabling \mmdb to achieve lower response times across all tasks.
(2) As shown in Figure~\ref{fig: ablation graph}, for single-hop graph queries such as G1 and G2, the performance gains achieved by using an adjacency-list-based DFS/BFS operator alone are limited. This is because a significant portion of the execution time is still spent on scanning large record tables to evaluate attribute predicates. In contrast, \mmdb employs topology- and attribute-aware graph operators that enable predicate pushdown whenever possible, thereby substantially reducing unnecessary table scans and improving overall query efficiency.
(3) With larger data, the acceleration effect of operators combined with optimization mechanisms becomes more evident (see G2 in Figure~\ref{fig: ablation gcq}), since \bench includes only \ti with simple topologies and the optimization in \mmdbd (depending on the number of joined collections) is limited. In contrast, the query optimizations in Section~\ref{sec: query optimization} can filter redundant records early, yielding greater benefits with few but large collections.
(4) For some tasks, response time decreases as data grows, because \bench scales data randomly, introducing more mismatches in match and join operations, producing smaller intermediates and shorter execution times.

\input{runtime}

\begin{figure*}[t]
    \centering
    \includegraphics[width=\textwidth]{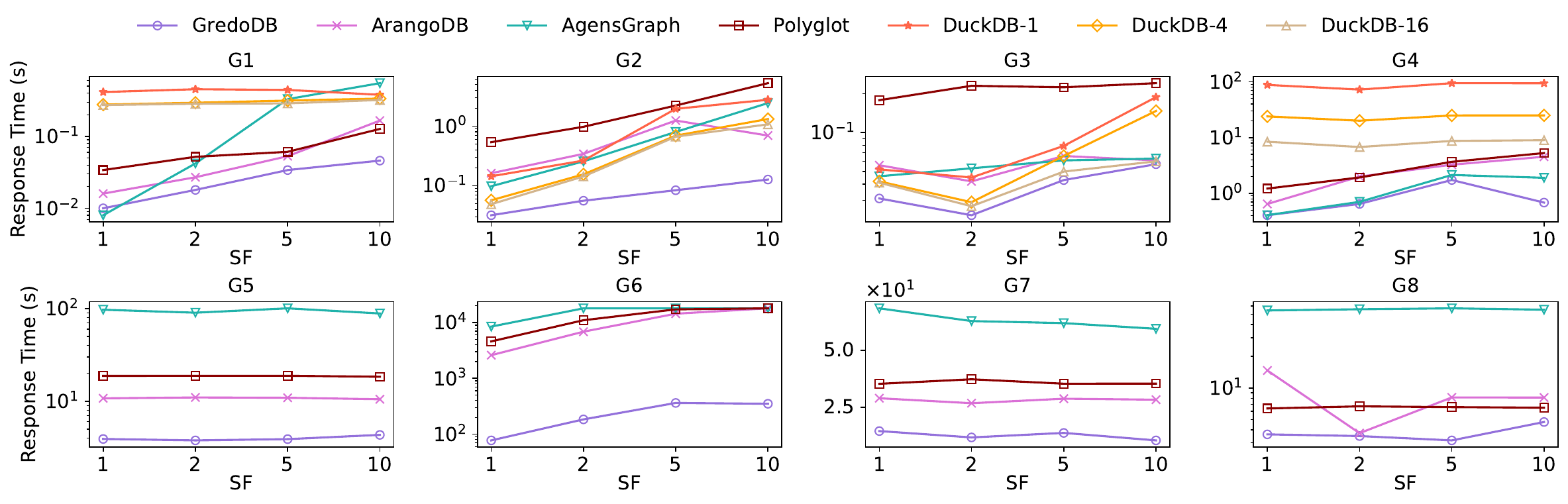}
    \caption{Evaluation on graph model workloads.}
    \label{fig: graph workload}
\end{figure*}

\myparagraph{Optimization Effectiveness for \ta}
For analytical tasks in \bench, A1-A3 involve local data access for \ta such as logistic regression and similarity computation, while A4-A6 rely on random access followed by basic aggregation. 
In this work, we design analytical operators specifically for local-access patterns and their downstream computations (see Section~\ref{sec: Analysis Operators}). Optimization for random-access workloads and aggregation-based analysis is out of the scope of our framework (as discussed in Section~\ref{sec: related work}). Accordingly, we evaluate all three variants on A1-A3, where our optimizations are applicable, and report the results in Figure~\ref{fig: ablation analysis}. The response times for A4-A6 are identical across all variants and therefore omitted.

\finding
\mmdb consistently achieves lower response times than \mmdbs and \mmdbd across A1-A3, indicating that the proposed parallel analytical pipeline significantly accelerates \ti compared with volcano-model-based execution.

\subsection{Comparison with SOTA Systems}
\label{sec: response time}

\myparagraph{Performance Overview on \bench}
Table~\ref{tab: m2ench-runtime} reports the response times of comparison systems at all different scale factors (SF = 1, 2, 5, and 10). We denote \duck with 1, 4, and 16 threads as \texttt{DuckDB-1}, \texttt{DuckDB-4}, and \texttt{DuckDB-16}.

\finding
(1) \mmdb outperforms other SOTA systems with a speedup of $11.73 \times$ to $34.72 \times$ in \texttt{SUM} and $1.73 \times$ to $4.07 \times$ in \texttt{GEOMEAN}, demonstrating markedly superior performance on \task tasks.
(2) \arango shows the best response times on four tasks, mainly because they involve large-scale document access (4.2 million to 28.2 million), where its document-first design and dedicated index optimizations are most effective.
(3) Although multi-threading improves \duck's query performance, it still exhibits limitations in multi-model query scenarios. This is primarily because its support for graph and document data relies on a plugin-based approach, without additional mechanisms for inter-model interaction or optimization. Furthermore, during spatial data import (t10-t16) and some query executions (t1 when SF = 2, 5, and 10), we observed kernel errors that prevented results from being obtained. We report the response times of these queries as N/A.

\begin{figure*}[t]
    \centering
    \includegraphics[width=\textwidth]{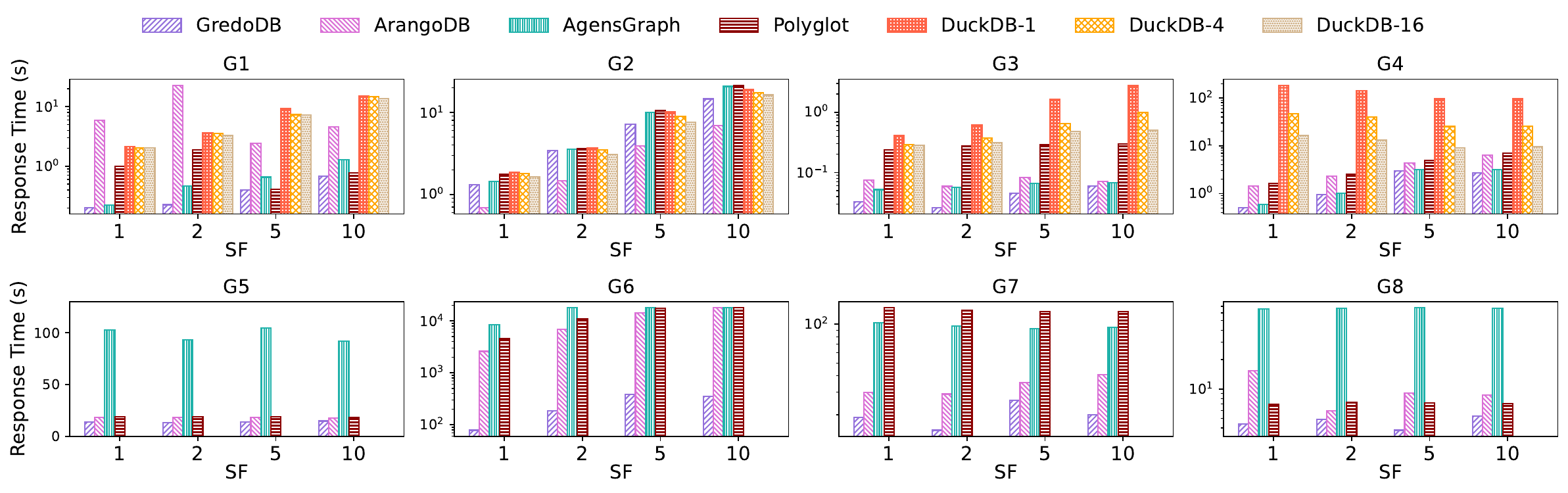}
    \caption{Evaluation of response time (seconds) on \ti (SF = 1, 2, 5, and 10).}
    \label{fig: gcq}
\end{figure*}

\begin{figure*}[t]
    \centering
    \includegraphics[width=\textwidth]{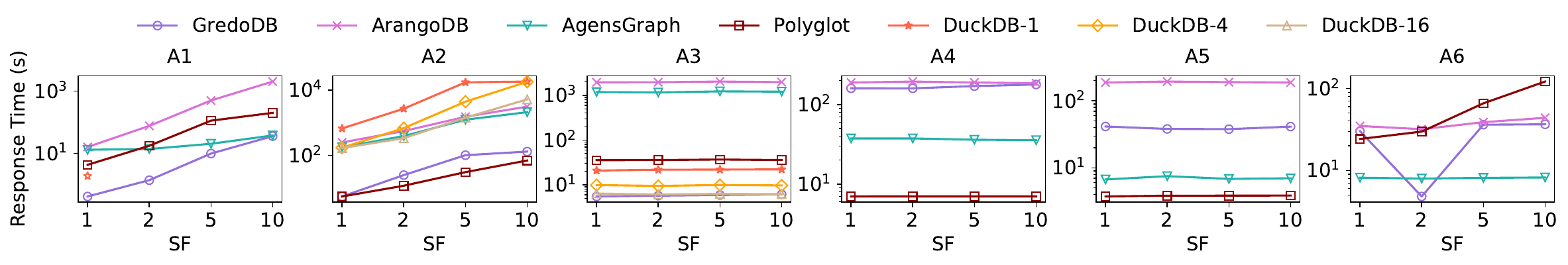}
    \caption{Evaluation of response time (seconds) on analysis tasks (SF = 1, 2, 5, and 10), where A1-A3 refer to \ta.}
    \label{fig: array-analysis}
\end{figure*}

\begin{figure*}[t]
    \centering
    \includegraphics[width=\textwidth]{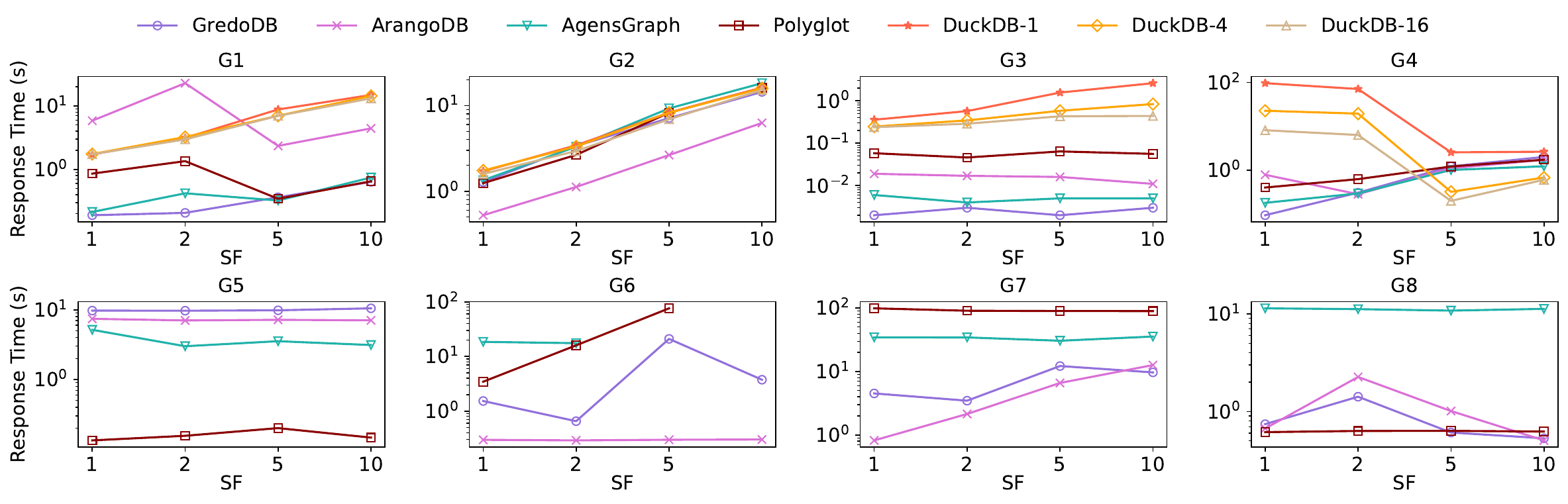}
    \caption{Evaluation on relational and document models.}
    \label{fig: rd workload}
\end{figure*}

\myparagraph{Response Time on Graph Processing}
We evaluate the query performance on graph workloads using G1-G8, and summarize the response times in Figure~\ref{fig: graph workload}.
Among these query tasks, G1-G5 mainly evaluate the efficiency of pattern matching, whereas G6-G8 focus on the shortest-path search performance.

\finding
(1) \mmdb outperforms all other comparison systems on graph workloads (except G1), consistently achieving lower response times.
(2) For pattern-matching workloads (G1-G5), compared to the hash indexes in \arango, the B-tree indexes in \agens, and the multi-way join-based implementation in \duck with \duckpgq, \mmdb’s graph-centric operators combined with topology- and attribute-aware optimizations deliver substantially higher efficiency.
(3) As a representative GNS, \arango generally outperforms systems based on other architectures (e.g., \poly under the MES paradigm and \agens under the TBS paradigm) on most graph queries. This advantage stems from its dedicated graph topology storage and traversal physical operators, which enable efficient graph exploration. In certain cases (e.g., G2 and G4), \arango exhibits longer response times than \agens. We attribute this behavior to the overhead incurred during graph traversal, where hash-index-based scans require frequent accesses to attribute records, leading to increased I/O costs. In contrast, \mmdb introduces explicit pruning mechanisms (as described in Section~\ref{sec: query optimization}) and relies on $tid$-based record access methods to reduce unnecessary attribute accesses, thereby effectively mitigating I/O overhead.
(4) \agens performs the worst on shortest-path workloads, as it lacks specialized operators and relies instead on costly multi-way joins for topology-only queries.
(5) Although \poly employs \texttt{Neo4j} for graph queries, its performance remains poor because the polyglot architecture incurs significant cross-engine communication overhead.

\myparagraph{Efficiency on \ti}
Figure~\ref{fig: gcq} illustrates the comparison of \ti using G1-G8. Note that response time is reported as the sum of relational, document, and graph queries rather than overall query time, which differs only when analytical tasks are included.

\finding
(1) \mmdb achieves superior response times over any other systems on all \ti (except G2), with up to a $101.45 \times$ speedup over \arango (on G1) and $107.89 \times$ over \agens (on G6), highlighting the impact of our operator, optimizer, and planner design in accelerating \ti execution. In G2, \mmdb exhibits suboptimal response time as a significant portion of execution is spent on processing document data, for which \mmdb lacks dedicated optimization support.
{
(2) \duck fails to execute G5-G8 because it lacks support for spatial data. Even on G1-G4, its performance is limited by the inefficiency of \duckpgq for graph queries and its JSON extension for document queries, as discussed in Section~\ref{sec: Single-model workloads}.}

\myparagraph{Efficiency on Analytical Workloads}
We summarize the performance of all systems on A1-A6 in Figure~\ref{fig: array-analysis} to evaluate their analysis capabilities. For \duck, kernel errors occurred on A1 (SF = 2, 5, and 10) and A4-A6, preventing execution results from being obtained.

\finding
(1) On A1-A3, \mmdb achieves lower response times than other systems, owing to its native parallel \ta processing strategy. On A3, it achieves up to a $356.72 \times$ speedup over \arango.
(2) \mmdb shows suboptimal response time on A4 and A6, as these random access tasks (A4-A6) involve simple aggregations executed by relational-style operators, which bypass the analytical pipeline and thus gain no benefit from its parallel optimization.
(3) \arango performs worst across most analytical tasks, primarily because it relies on volcano model execution and key-based data access. The former incurs overhead under the tuple-at-a-time paradigm, while the latter causes inefficient scans compared to the physical-address-based access in \mmdb and \agens.
(4) \poly delivers the best performance on many tasks, owing to \texttt{SciDB}’s specialized storage and execution engine optimized for chunk-based processing and multidimensional indexing. Nevertheless, on A1-A3, where \mmdb’s parallel optimization takes effect, \mmdb still delivers comparable or superior performance.

\myparagraph{Impact of Data Scale Factors}
We evaluate all systems on \ti tasks (G1-G8) and analytical tasks (A1-A6) under different scale factors, with the response times shown in Figure~\ref{fig: gcq} and Figure~\ref{fig: array-analysis}.

\finding 
(1) Across different scale factors, all systems maintain performance trends consistent with those observed under SF = 1. In particular, \mmdb consistently outperforms the other systems in all \ti tasks (except G2) and in most complex \ta workloads (A1-A3), demonstrating stable and robust scalability.
(2) As data scale increases, several systems encounter query timeouts, e.g., \arango, \agens, and \poly in G6 (SF=10), and \texttt{DuckDB-1} in A2 (SF=10), due to the lack of dedicated optimizations for \task workloads. In contrast, \mmdb consistently maintains lower response times across all scale factors.

\subsection{Performance on Individual Workloads} \label{sec: Single-model workloads}
Since the execution time of \ti is the sum of the processing times over relational, document, and graph models, we design a single-model workload evaluation to investigate which model has the greatest impact in each system. To this end, we analyze the query execution plans and measure the response times of the physical operators corresponding to relational and document models. The graph model is discussed in Section~\ref{sec: response time}.

\myparagraph{Performance on Document and Relational Workloads}
{
We report the total response time of document and relational operators for G1-G8 in Figure~\ref{fig: rd workload}. Relational operators dominate in G1, G4, G5, G7, whereas document operators dominate in G2, G3, G6, G8.

\finding
(1) Although no dedicated relational or document optimizations are introduced, \mmdb still demonstrates overall competitive performance. This improvement primarily results from our \ti optimizations, which generate lower-cost execution plans and consequently reduce relational and document processing time.
(2) With its document-first design, \arango performs best on most document tasks, suggesting that its primary optimization objectives for \ti should lie in relational and graph queries.
(3) Despite the benefits of multi-threaded acceleration, \duck shows poor performance on all \ti. The plugin-based multi-model design restricts data interaction and optimizations, and the lack of JSON indexing further makes it unsuitable for \ti workloads.}

%% file: model.tex
\begin{table}[t]\small
\setlength{\tabcolsep}{0.5pt}
\ra{1.05}
\centering
\caption{Workloads involved in the query tasks of \bench (K denotes thousand, M denotes million).}
\label{tab: model}
\begin{tabular}{cccc}
\toprule
\textbf{Workloads}      & \textbf{Query Tasks}     & \textbf{Alias}  & \textbf{Data Scales}    \\ \midrule
Relational  & t0-t3, t7, t10, t12        & R1-R7  & 19.7K-84.2M \\
Document    & t0-t6, t9-t14, t16  & D1-D14 & 17.4K-28.2M \\
Graph       & t3-t5, t7, t10-t12, t15  & G1-G8  & 0.8M-69.8M edges \\
Analysis    & t0, t2, t9, t14-t16           & A1-A6 & N/A \\

\bottomrule
\end{tabular}
\end{table}

%% file: runtime.tex
\begin{table*}[t]\small
\setlength{\tabcolsep}{0.72pt}
\centering
\ra{1.04}
\definecolor{softteal}{RGB}{150,120,240}
\caption{Comparison of different systems in terms of response time (seconds) on M2Bench (SF = 1, 2, 5 and 10). The best and second-best results for each query are highlighted in dark and light purple. Queries exceeding the time limit are marked TLE and assigned $1.8 \times 10^4$ seconds, while unexecutable ones are denoted N/A. \texttt{SUM} and \texttt{GEOMEAN} represent the total and geometric mean of response times, respectively.
}
\label{tab: m2ench-runtime}

\begin{tabular}{ccccccccccccccccccccc}
\toprule

\multirow{2}{*}{\textbf{\rule{0pt}{3.5ex}SF}} & \multirow{2}{*}{\textbf{\rule{0pt}{3.5ex}MMDBs}} & \multicolumn{17}{c}{\textbf{Query Tasks}} & \multirow{2}{*}{\textbf{\rule{0pt}{3.5ex}\texttt{SUM}}} & \multirow{2}{*}{\textbf{\rule{0pt}{3.5ex}\texttt{GEOMEAN}}} \\ \cmidrule{3-19}

&       
& t0 & t1 & t2 & t3 & t4 
& t5 & t6 & t7 & t8 
& t9 & t10 & t11 & t12 
& t13& t14 & t15 & t16 &                  \\ \midrule

\multirow{7}{*}{1}      
& \mmdb
& \cellcolor{softteal!50}{1.713} & 0.369 & \cellcolor{softteal!50}{7.198} & \cellcolor{softteal!50}{0.200} & \cellcolor{softteal!20}{{1.307}}
& \cellcolor{softteal!50}{0.033} & 0.542 & \cellcolor{softteal!50}{0.501} & \cellcolor{softteal!50}{0.168}
& \cellcolor{softteal!20}{9.213} & \cellcolor{softteal!50}{13.63} & \cellcolor{softteal!50}{78.68} & \cellcolor{softteal!50}{19.01} 
& 36.41 & 160.6 & \cellcolor{softteal!20}{45.82} & 44.19 & \cellcolor{softteal!50}{419.5} & \cellcolor{softteal!50}{3.671} \\

& \arango
& 190.9 & \cellcolor{softteal!50}{0.110} & 267.2 & 5.862 & \cellcolor{softteal!50}{0.686}
& 0.075 & \cellcolor{softteal!50}{0.414} & 1.438 & \cellcolor{softteal!20}{0.888}
& 1958  & \cellcolor{softteal!20}{18.16} & \cellcolor{softteal!20}{2603}  & \cellcolor{softteal!20}{29.73} 
& \cellcolor{softteal!50}{14.32} & 190.1 & 203.2 & 51.57 & 5536 & {14.71} \\

& \agens
& 19.66 & 0.155 & 179.9 & \cellcolor{softteal!20}{0.222} & 1.443
& \cellcolor{softteal!20}{0.052} & 0.580 & \cellcolor{softteal!20}{0.590} & 4.433 
& 1188  & 102.3 & 8489  & 102.8 
& 41.63 & \cellcolor{softteal!20}{37.64} & 72.55 & \cellcolor{softteal!50}{15.69} & 10257 & {12.12} \\ 

& \poly
& 13.48 & \cellcolor{softteal!20}{0.130} & \cellcolor{softteal!20}{19.03} & 0.893 & 1.778
& 0.236 & \cellcolor{softteal!20}{0.477} & 1.625 & 27.58 & 36.53
& 18.91 & 4601  & 133.9 & \cellcolor{softteal!20}{15.37} & \cellcolor{softteal!50}{6.964}
& \cellcolor{softteal!50}{10.73} & \cellcolor{softteal!20}{34.03} & \cellcolor{softteal!20}{4923} & \cellcolor{softteal!20}{8.380} \\

& \texttt{DuckDB-1}
& 2.424 & 2.837 & 664.6 & 2.133 & 1.851
& 0.410 & 1.037 & 184.1 & 7.109 & 24.12
& {\color{black} N/A}   & {\color{black} N/A}   & {\color{black} N/A}   & {\color{black} N/A}   
& {\color{black} N/A}   & {\color{black} N/A}   & {\color{black} N/A}   & {\color{black} N/A} & {\color{black} N/A} \\ 

& \texttt{DuckDB-4}
& \cellcolor{softteal!20}{2.249} & 0.967 & 171.5 & 2.022 & 1.800
& 0.289 & 0.992 & 46.64 & 7.006 & 12.38
& {\color{black} N/A}   & {\color{black} N/A}   & {\color{black} N/A}   & {\color{black} N/A}   
& {\color{black} N/A}   & {\color{black} N/A}   & {\color{black} N/A}   & {\color{black} N/A} & {\color{black} N/A} \\ 

& \texttt{DuckDB-16}
& 2.338 & 0.513 & 167.3 & 2.006 & 1.640
& 0.280 & 0.977 & 16.61 & 6.225 & \cellcolor{softteal!50}{8.259}
& {\color{black} N/A}   & {\color{black} N/A}   & {\color{black} N/A}   & {\color{black} N/A}   
& {\color{black} N/A}   & {\color{black} N/A}   & {\color{black} N/A}   & {\color{black} N/A} & {\color{black} N/A} \\ \midrule

\multirow{7}{*}{2}      
& \mmdb
&\cellcolor{softteal!50}{6.420}&	0.646&	\cellcolor{softteal!50}{27.94}&	\cellcolor{softteal!50}{0.225}	&3.428&	\cellcolor{softteal!50}{0.026}&	\cellcolor{softteal!20}{0.601}&	\cellcolor{softteal!50}{0.957}&	\cellcolor{softteal!50}{0.025}&	\cellcolor{softteal!20}{9.205}	&\cellcolor{softteal!50}{13.46}	&\cellcolor{softteal!50}{184.2}	&\cellcolor{softteal!50}{15.19}	&35.78	&160.2	&\cellcolor{softteal!20}{43.01}	&49.66 & \cellcolor{softteal!50}{551.1} & \cellcolor{softteal!50}{4.529}
 \\

& \arango
& 789.1& 	\cellcolor{softteal!50}{0.408}& 	559.1& 	22.82& 	\cellcolor{softteal!50}{1.463}& 	0.059& 	\cellcolor{softteal!50}{0.361}& 	2.286	
& \cellcolor{softteal!20}{0.945}& 	1972& 	\cellcolor{softteal!20}{18.01}& 	\cellcolor{softteal!20}{6863}& 	\cellcolor{softteal!20}{28.87}& 	\cellcolor{softteal!50}{13.19}& 	198.8& 	200.2& 	63.42 & \cellcolor{softteal!20}{10735}  & 21.97
 \\

& \agens
& \cellcolor{softteal!20}{38.54}&	\cellcolor{softteal!20}{0.213}&	392.5&	\cellcolor{softteal!20}{0.463}&	3.562&	\cellcolor{softteal!20}{0.057}&	0.621&	\cellcolor{softteal!20}{1.003}&	4.925
&	1166&	93.35&	{\color{black} TLE}&	97.14&	40.99&	\cellcolor{softteal!20}{37.49}&	74.61&	\cellcolor{softteal!50}{15.34} & 19967  & 16.05
 \\ 

& \poly
& 54.58&	0.788&	\cellcolor{softteal!20}{39.84}&	1.402&	3.593&	0.276&	0.693&	2.560&	29.35
&	36.98&	19.00&	11032&	127.7&	\cellcolor{softteal!20}{14.15}&	\cellcolor{softteal!50}{6.971}&	\cellcolor{softteal!50}{11.16}&	\cellcolor{softteal!20}{41.85} & 11423  & \cellcolor{softteal!20}{12.68}
 \\

& \texttt{DuckDB-1}
& {\color{black} N/A}	&5.230	&2669	&3.589	&3.705	&0.617	&1.178	&143.3	&6.784	&25.52
& {\color{black} N/A}   & {\color{black} N/A}   & {\color{black} N/A}   & {\color{black} N/A}   
& {\color{black} N/A}   & {\color{black} N/A}   & {\color{black} N/A}   & {\color{black} N/A} & {\color{black} N/A} \\ 

& \texttt{DuckDB-4}
& {\color{black} N/A}	&1.559	&679.7	&3.534	&3.486	&0.374	&1.094	&39.55	&6.318	&11.84
& {\color{black} N/A}   & {\color{black} N/A}   & {\color{black} N/A}   & {\color{black} N/A}   
& {\color{black} N/A}   & {\color{black} N/A}   & {\color{black} N/A}   & {\color{black} N/A} & {\color{black} N/A} \\ 

& \texttt{DuckDB-16}
& {\color{black} N/A}	&0.889	&334.9	&3.256	&\cellcolor{softteal!20}{3.092}	&0.315	&0.993	&13.14	&6.193	&\cellcolor{softteal!50}{7.804}
& {\color{black} N/A}   & {\color{black} N/A}   & {\color{black} N/A}   & {\color{black} N/A}   
& {\color{black} N/A}   & {\color{black} N/A}   & {\color{black} N/A}   & {\color{black} N/A} & {\color{black} N/A} \\ \midrule

\multirow{7}{*}{5}      
& \mmdb
& \cellcolor{softteal!50}{44.42}&	1.690&	\cellcolor{softteal!50}{109.4}&	\cellcolor{softteal!50}{0.398}&	\cellcolor{softteal!20}{7.215}&	\cellcolor{softteal!50}{0.045}&	0.633&	\cellcolor{softteal!50}{2.969}&	\cellcolor{softteal!50}{0.016}&	\cellcolor{softteal!50}{3.463}&	\cellcolor{softteal!50}{13.71}&	\cellcolor{softteal!50}{383.1}&	\cellcolor{softteal!50}{25.87}&	35.26&	171.2&	\cellcolor{softteal!20}{41.60}&	\cellcolor{softteal!20}{51.68}&	\cellcolor{softteal!50}{892.9}&	\cellcolor{softteal!50}{6.928}
 \\

& \arango
& 5058&	\cellcolor{softteal!20}{0.696}&	1646&	2.401&	\cellcolor{softteal!50}{3.881}&	0.082&	\cellcolor{softteal!50}{0.521}&	4.374&	\cellcolor{softteal!20}{0.018}&	2040&	\cellcolor{softteal!20}{18.06}&	\cellcolor{softteal!20}{14415}&	\cellcolor{softteal!20}{35.30}&	\cellcolor{softteal!50}{13.13}&	195.2&	199.6&	140.2&	23775&	23.75
 \\

& \agens
& \cellcolor{softteal!20}{187.1}&	\cellcolor{softteal!50}{0.560}&	1242&	0.654&	10.06&	\cellcolor{softteal!20}{0.066}&	0.664&	\cellcolor{softteal!20}{3.156}&	5.406&	1233&	104.2&	{\color{black} TLE}&	92.34&	42.01&	\cellcolor{softteal!20}{36.11}&	74.69&	\cellcolor{softteal!50}{15.39}&	\cellcolor{softteal!20}{3048}&	15.67
 \\ 

& \poly
& 350.7&	0.849&	\cellcolor{softteal!20}{117.3}&	\cellcolor{softteal!20}{0.406}&	10.47&	0.288&	\cellcolor{softteal!20}{0.599}&	4.899&	0.559&	38.07&	19.03&	17325&	124.7&	\cellcolor{softteal!20}{14.09}&	\cellcolor{softteal!50}{6.981}&	\cellcolor{softteal!50}{11.05}&	92.53&	18118&	\cellcolor{softteal!20}{13.18}
 \\

& \texttt{DuckDB-1}
& {\color{black} N/A}&	14.04&	17040&	9.227&	10.22&	1.644&	1.017&	96.86&	6.946&	25.55
& {\color{black} N/A}   & {\color{black} N/A}   & {\color{black} N/A}   & {\color{black} N/A}   
& {\color{black} N/A}   & {\color{black} N/A}   & {\color{black} N/A}   & {\color{black} N/A} & {\color{black} N/A} \\ 

& \texttt{DuckDB-4}
& {\color{black} N/A}&	4.449&	4377&	7.314&	8.934&	0.648&	0.979&	25.22&	6.819&	11.93
& {\color{black} N/A}   & {\color{black} N/A}   & {\color{black} N/A}   & {\color{black} N/A}   
& {\color{black} N/A}   & {\color{black} N/A}   & {\color{black} N/A}   & {\color{black} N/A} & {\color{black} N/A} \\ 

& \texttt{DuckDB-16}
& {\color{black} N/A}&	2.097&	1412&	7.268&	7.577&	0.482&	0.954&	8.907&	6.623&	\cellcolor{softteal!20}{8.011}
& {\color{black} N/A}   & {\color{black} N/A}   & {\color{black} N/A}   & {\color{black} N/A}   
& {\color{black} N/A}   & {\color{black} N/A}   & {\color{black} N/A}   & {\color{black} N/A} & {\color{black} N/A} \\ \midrule

\multirow{7}{*}{10}      
& \mmdb
& \cellcolor{softteal!50}{332.1} & 3.083 & \cellcolor{softteal!50}{148.4} & \cellcolor{softteal!50}{0.684} & \cellcolor{softteal!20}{14.57} 
& \cellcolor{softteal!50}{0.060} & 0.622 & \cellcolor{softteal!50}{2.702} & \cellcolor{softteal!50}{0.086} & \cellcolor{softteal!50}{3.863}
& \cellcolor{softteal!50}{14.81} & \cellcolor{softteal!50}{353.8} & \cellcolor{softteal!50}{20.09} & 36.39 & 179.3
& \cellcolor{softteal!20}{46.54} & \cellcolor{softteal!20}{53.79} & \cellcolor{softteal!50}{1211} & \cellcolor{softteal!50}{9.920} \\

& \arango
& {\color{black} TLE}	& \cellcolor{softteal!20}{2.252}	& 3340	& 4.591	& \cellcolor{softteal!50}{{6.963}}	
& 0.072	& \cellcolor{softteal!50}{0.430}	& 6.271	& {0.800}	& 1981	
& \cellcolor{softteal!20}{17.55}	& {\color{black} TLE}	& \cellcolor{softteal!20}{40.88}	& \cellcolor{softteal!50}{12.45}	& 188.2	
& 196.5	& 258.2	& 42056 & 40.41 \\

& \agens
& \cellcolor{softteal!20}{653.9}	& \cellcolor{softteal!50}{0.988}	& 21139	& 1.296	& 20.69	
& \cellcolor{softteal!20}{0.068}	& 0.683	& \cellcolor{softteal!20}{3.136}	& 4.793	& 1216	
& 91.74	& {\color{black} TLE}	& 94.87	& 40.60	& \cellcolor{softteal!20}{35.57}
& 73.61	& \cellcolor{softteal!50}{15.33}	& 22367 & 29.16 \\ 

& \poly
& 704.3	& 2.743	& \cellcolor{softteal!20}{235.2}	& \cellcolor{softteal!20}{0.778}	& 21.39
& 0.297	& \cellcolor{softteal!20}{0.498}	& 7.024	& \cellcolor{softteal!20}{0.496}	& 36.96
& 18.52	& {\color{black} TLE}	& 124.6	& \cellcolor{softteal!20}{13.36}	& \cellcolor{softteal!50}{6.982}
& \cellcolor{softteal!50}{10.97}	& 170.4	& \cellcolor{softteal!20}{19354} & \cellcolor{softteal!20}{17.24}\\

& \texttt{DuckDB-1}
& {\color{black} N/A}   & 25.73 & {\color{black} TLE}   & 15.26 & 18.96
& 2.795 & 0.960 & 96.55 & 6.880 & 25.69
& {\color{black} N/A}   & {\color{black} N/A}   & {\color{black} N/A}   & {\color{black} N/A}   
& {\color{black} N/A}   & {\color{black} N/A}   & {\color{black} N/A}   & {\color{black} N/A} & {\color{black} N/A}  \\ 

& \texttt{DuckDB-4}
& {\color{black} N/A}   & 9.400 & 17619 & 14.70 & 17.24
& 0.985 & 0.940 & 25.67 & 6.451 & 11.69
& {\color{black} N/A}   & {\color{black} N/A}   & {\color{black} N/A}   & {\color{black} N/A}   
& {\color{black} N/A}   & {\color{black} N/A}   & {\color{black} N/A}   & {\color{black} N/A} & {\color{black} N/A}  \\ 

& \texttt{DuckDB-16}
& {\color{black} N/A}   & 4.763 & 5140  & 13.48 & 16.29
& 0.500 & 0.938 & 9.581 & 5.843 & \cellcolor{softteal!20}{7.86}
& {\color{black} N/A}   & {\color{black} N/A}   & {\color{black} N/A}   & {\color{black} N/A}   
& {\color{black} N/A}   & {\color{black} N/A}   & {\color{black} N/A}   & {\color{black} N/A} & {\color{black} N/A} \\

\bottomrule
\end{tabular}

\end{table*}